\documentclass[twocolumn]{aastex62}

\usepackage{graphicx}	
\usepackage{amsmath}	
\usepackage{amssymb}	
\usepackage{float}
\usepackage{hyperref}
\usepackage{algorithm}
\usepackage[noend]{algpseudocode}
\usepackage[FIGTOPCAP]{subfigure}
\usepackage{tabularx}

\def\ln{\mathrm{ln}}
\def\log10{\mathrm{log}_{10}}

\def\flux{\mathbf{f}}

\def\sps{\boldsymbol{\varphi}}
\def\appearance{\boldsymbol{\zeta}}
\def\observe{\mathcal{O}}
\def\data{\mathbf{d}}
\def\Data{\mathbf{D}}
\def\noise{\boldsymbol{\sigma}}
\def\Noise{\boldsymbol{\Sigma}}
\def\hyper{\boldsymbol{\psi}}
\def\nuisance{\boldsymbol{\eta}}

\graphicspath{{./}{figures/}}

\received{}
\revised{}
\accepted{}
\submitjournal{ApJS}

\shorttitle{Forward modeling for redshift distribution inference}
\shortauthors{Alsing et al.}

\begin{document}

\title{Forward modeling of galaxy populations for cosmological redshift distribution inference}

\correspondingauthor{Justin Alsing}
\email{justin.alsing@fysik.su.se}

\author[0000-0003-4618-3546]{Justin Alsing}
\affiliation{Oskar Klein Centre for Cosmoparticle Physics, Department of Physics, Stockholm University, Stockholm SE-106 91, Sweden}

\author[0000-0002-2519-584X]{Hiranya Peiris}
\affiliation{Department of Physics and Astronomy, University College London, Gower Street, London, WC1E 6BT, UK}
\affiliation{Oskar Klein Centre for Cosmoparticle Physics, Department of Physics, Stockholm University, Stockholm SE-106 91, Sweden}

\author[0000-0002-0041-3783]{Daniel Mortlock}
\affiliation{Department of Physics, Imperial College London, Blackett Laboratory, Prince Consort
Road, London SW7 2AZ, UK}
\affiliation{Department of Mathematics, Imperial College London, London SW7 2AZ, UK}
\affiliation{Oskar Klein Centre for Cosmoparticle Physics, Department of Physics, Stockholm University, Stockholm SE-106 91, Sweden}

\author[0000-0001-6755-1315]{Joel Leja}
\affil{Department of Astronomy \& Astrophysics, The Pennsylvania State University, University Park, PA 16802, USA}
\affil{Institute for Computational \& Data Sciences, The Pennsylvania State University, University Park, PA, USA}
\affil{Institute for Gravitation and the Cosmos, The Pennsylvania State University, University Park, PA 16802, USA}

\author[0000-0002-3962-9274]{Boris Leistedt}
\affiliation{Department of Physics, Imperial College London, Blackett Laboratory, Prince Consort
Road, London SW7 2AZ, UK}


\begin{abstract}
We present a forward modeling framework for estimating galaxy redshift distributions from photometric surveys. Our forward model is composed of: a detailed population model describing the intrinsic distribution of physical characteristics of galaxies, encoding galaxy evolution physics; a stellar population synthesis model connecting the physical properties of galaxies to their photometry; a data-model characterizing the observation and calibration processes for a given survey; and, explicit treatment of selection cuts, both into the main analysis sample and subsequent sorting into tomographic redshift bins. This approach has the appeal that it does not rely on spectroscopic calibration data, provides explicit control over modeling assumptions, and builds a direct bridge between photo-$z$ inference and galaxy evolution physics. In addition to redshift distributions, forward modeling provides a framework for drawing robust inferences about the statistical properties of the galaxy population more generally. We demonstrate the utility of forward modeling by estimating the redshift distributions for the Galaxy And Mass Assembly (GAMA) and Vimos VLT Deep (VVDS) surveys, validating against their spectroscopic redshifts. Our baseline model is able to predict tomographic redshift distributions for GAMA and VVDS with a bias of $\Delta z \lesssim 0.003$ and $\Delta z \simeq 0.01$ on the mean redshift respectively -- comfortably accurate enough for Stage III cosmological surveys -- \emph{without any hyper-parameter tuning} (i.e., prior to doing any fitting to those data). We anticipate that with additional hyper-parameter fitting and modeling improvements, forward modeling can provide a path to accurate redshift distribution inference for Stage IV surveys.
\end{abstract}

\keywords{photometric redshifts - galaxy surveys - cosmological parameters}

\section{Introduction}
Accurate inference of the redshift distributions of ensembles of galaxies from their photometry is of central importance for deriving cosmological constraints from weak lensing surveys. Ongoing and upcoming surveys such as the Dark Energy Survey \citep{des}, the Kilo-Degree Survey \citep{kids}, Hyper Suprime-Cam \citep{hsc}, the Vera C. Rubin Observatory's Legacy Survey of Space and Time \citep{lsst} and \emph{Euclid} \citep{euclid} will map large volumes of the Universe, measuring the angular positions, images, and photometry for billions of galaxies. The unprecedented statistical power of these surveys makes them increasingly sensitive to systematic biases, with systematics on the redshift distributions in particular expected to be the single largest contributor to the total systematic error budget (see e.g., \citealp{hildebrandt2017}). In order to reach the full potential of upcoming Stage IV surveys, the characterization of the redshift distributions (e.g., their measured means and variances) will need to improve by roughly an order of magnitude compared to the current state-of-the-art \citep{newman2022}.

Three main approaches exist for estimating cosmological redshift distributions: cross-correlation \citep{schneider2006, newman2008, mcquinn2013, menard2013, schmidt2013, morrison2017, davis2018}, direct calibration \citep{lima2008, hildebrandt2016, hildebrandt2020, buchs2019, wright2020}, and template-based methods \citep{benitez2000, ilbert2006, brammer2008, arnouts2011, hildebrandt2012, hoyle2018, tanaka2018, leistedt2016, leistedt2019}.

Cross-correlation and direct calibration both rely on spectroscopic redshift samples to compare against the photometric data in order to calibrate their redshift distributions. These approaches have the appeal that they leverage reliably measured redshifts to calibrate the photo-$z$ distributions, and are relatively insensitive to modeling assumptions about the photometric data. On the other hand, they are limited by the lack of available spectroscopic redshifts at the depths probed by ongoing and upcoming surveys, and are vulnerable to biases due to spectroscopic selection effects that are not well-represented by re-weighting in the (broad-band) colors \citep{buchs2019, hartley2020}, and uncertainties in galaxy-bias modeling in the case of cross-correlation methods \citep{gatti2018}.

Template-based methods instead rely on building a statistical model for the photometric observations in order to constrain the redshifts of individual galaxies, and the redshift distributions of ensembles, from their photometry alone. Template approaches assert that galaxies belong to one of a finite set of ``types", where each type has an associated (rest-frame) spectral template that defines its colors as a function of redshift. These templates can then be compared to the observed colors to give redshift (and type) likelihoods for each galaxy, which can then be combined in a hierarchical model to infer the redshift distributions of ensembles of galaxies \citep{leistedt2016, leistedt2019}. However, template methods are limited by the incapability of template-sets (and priors over galaxy types) to characterize the galaxy population at the necessary level of realism. Finite template sets tend to be overly restrictive, while methods that allow for linear combinations of templates result in large swathes of prior volume populated by unphysical galaxy spectra. Further, by relying on a model for the photometric data, template-based models need to treat selection effects explicitly in order to draw robust inferences at the population level (e.g., redshift distributions): this has so far not been done.

Physically, the redshift distributions of selected samples of galaxies arise from a sequence of three main processes: The statistical properties of the galaxy population (i.e., the intrinsic distribution of physical characteristics and redshifts) defines an intrinsic distribution for galaxy colors, from which galaxies in the Universe are sampled. The colors (photometry) of those galaxies in some patch of the sky then get observed by a survey, resulting in a catalog of noisy (measured) photometry. Selection cuts are then applied to those measurements to ensure a clean and high-quality galaxy sample, and to sort the galaxies into tomographic redshift bins. The redshift distributions of interest, then, are those of the galaxies that make it past the selection cuts and into a given tomographic bin. Therefore, if one is able to accurately characterize the galaxy population, observational processes, and selection effects, then one can predict the redshift distributions of interest.

In this paper we develop a forward modeling framework for estimating redshift distributions by explicitly modeling the processes that give rise to them. We construct a population model describing the joint distribution of physical characteristics (e.g., stellar, dust and gas content) of galaxies, encoding galaxy evolution physics in the relationships between galaxies physical properties. We use a stellar population synthesis (SPS) model to connect those physical parameters to the rest-frame spectra and hence photometry for each galaxy. The observation process is then characterized by a data-model that captures measurement noise, heterogeneous observing conditions and strategy, and photometric calibration. Finally, we have a selection model specifying the selection cuts. This parameterized forward model then forms the basis for Bayesian inference of cosmological redshift distributions, either by hierarchical inference, or simulation-based inference (SBI).

This forward modeling approach can be thought of as resolving the current limitations of template-based methods by replacing template-sets with a continuous SPS model, explicitly treating selection effects, and inferring population- and data-model parameters in a self-consistent fashion. In particular, in replacing finite template-sets by a continuous model we are able to better capture the diversity of real galaxy spectra whilst having full control over the prior describing the statistical properties of the galaxy population. The use of SPS models for analyzing large samples of galaxies has only recently become feasible, thanks to fast neural emulators (e.g., \texttt{speculator}, \citealp{alsing2020}). The use of physically motivated priors (e.g., \citealp{tanaka2015} and \citealp{ramachandra2021}) and continuous physical models for galaxy spectra \citep{ramachandra2021} has already led to promising improvements in photometric redshift inference for individual galaxies.

The structure of this paper is as follows. In \S \ref{sec:framework} we describe the framework for forward modeling photometric surveys, and estimating redshift distributions in a forward modeling context. In \S \ref{sec:forward_model} we describe our SPS model, galaxy population model, and data model assumptions. In \S \ref{sec:GAMA} and \ref{sec:VVDS} we show the ability of our baseline forward model to recover the tomographic redshift distributions for GAMA and VVDS respectively. We outline a roadmap for future forward modeling efforts and conclude in \S \ref{sec:conclusions}. 

In a companion paper \citep{leistedt2022} we validate the forward modeling framework for inferring individual galaxy redshifts, including hierarchical calibration of hyper-parameters.
\section{Forward modeling of galaxy surveys for redshift inference}
\label{sec:framework}
In this section we describe our generative modeling framework for photometric surveys. We frame the forward model as a pipeline for simulating mock catalogs, which can then be compared to the observed catalog in a simulation-based inference setting, used to estimate the implied tomographic redshift distributions for a given set of modeling and hyper-parameter choices, or can be used as a basis for hierarchical inference via MCMC sampling.

Notation is summarized in \S \ref{sec:notation} and Table \ref{tab:parameters}. We describe the generative model in \S \ref{sec:generate}, and discuss how to perform inference of the model parameters in \S \ref{sec:inference}.
\subsection{Notation}
\label{sec:notation}
Each galaxy is described by a set of SPS parameters $\sps$, which describe the stellar, gas and dust contents of the galaxy. The rest-frame spectrum $l(\lambda) \equiv l(\lambda ;\sps)$ is connected to the parameters $\sps$ via a stellar population synthesis model, which computes the composite spectrum from the stars in the population given their initial mass function, ages and metallicities from the star formation and metallicity histories, plus modifications due to dust, and nebular emission (see \citealp{conroy2013} for a review).

Combined with redshifts $z$, the SPS parameters predict the model photometry for each galaxy, i.e., the fluxes $\{f_b\}$ in band-passes $\{W_b(\lambda)\}$, defined by:
\begin{align}
    f_b&(\sps, z) = \nonumber \\
    & \frac{(1+z)^{-1}}{4\pi d_L^2(z)}\int_0^\infty l(\lambda/(1+z); \sps)e^{-\tau(z, \lambda)}W_b(\lambda)d\lambda,
\end{align}
where $d_L(z)$ the luminosity distance for redshift $z$ and $\tau(z, \lambda)$ is the optical depth of the inter-galactic medium.

We denote the vector of measured fluxes for each galaxy by $\data$, where the photometric \emph{data-model} specifies the sampling distribution $P(\data | \sps, z, \noise, \nuisance)$ of measured fluxes given the model fluxes and measurement uncertainties $\noise$. The data model is parameterized by nuisance parameters $\nuisance$, which characterize the properties of the noise distribution, calibration (e.g., zero-point) parameters, and modeling error terms.

The intrinsic distribution of galaxy characteristics (SPS parameters and redshift) is described by a \emph{population model} $P(\sps, z | \hyper)$, with hyper-parameters $\hyper$. This describes the joint distribution of galaxy characteristics for the background galaxy population, in the absence of any selection effects.

We assume that selection cuts for the main galaxy sample are made on observed photometry, with selection probability $P(S | \data, \noise)$ equal to one or zero for selection or rejection., i.e., selection is deterministic given the photometric data vector.

Subsequent sorting of galaxies into tomographic bins introduces an additional selection $S^{(k)}$ (for the $k$-th bin) based on the measured galaxy colors (fluxes). Typically, tomographic binning will be based on some estimator for the redshift, which will be a deterministic function of the measured fluxes and their uncertainties:
\begin{equation}
P(S^{(k)} | \hat{z}(\data, \noise) ) = \begin{cases}
1 & z^{(k)}_\mathrm{l} < \hat{z}(\data, \noise) < z^{(k)}_\mathrm{u}\\
0 &\text{otherwise}
\end{cases}
\end{equation}
where $\hat{z}(\data, \noise)$ is an estimator for the redshift, and $z^{(k)}_\mathrm{l}$ and $z^{(k)}_\mathrm{u}$ and the lower and upper limits for the $k$-th tomographic bin.

The measurement uncertainties on the photometry will be depend on the observing conditions (e.g., seeing), and strategy (e.g., exposure time), which will typically vary to some extent across the survey. The uncertainties will also scale with flux, owing to the Poisson photon count contribution to the errors, and have additional intrinsic scatter due to the varying difficulty in extracting fluxes from galaxy images with different morphologies. We denote the distribution of photometric uncertainties across the survey (as a function of flux) by the \emph{uncertainty model} $P(\noise | \flux(\sps, z), \observe)$, where $\observe$ denotes the model assumptions (and parameters) characterizing that distribution.
\begin{table*}
\centering
\scalebox{0.95}{
\begin{tabularx}{\textwidth}{cc}
\toprule
Parameter & Description \tabularnewline
\hline \tabularnewline
& \emph{Population-model parameters} \tabularnewline

$\hyper$ & Hyper-parameters describing the galaxy population model \tabularnewline

$\Phi_0$ & Present-day comoving volume-density of galaxies \tabularnewline

$\rho(z ; \hyper)$ & Evolution in the relative comoving number-density of galaxies ($\rho(0;\hyper) = 1$)\tabularnewline \tabularnewline

& \emph{Data-model parameters} \tabularnewline

$\boldsymbol{\eta}$ & Nuisance parameters determining the data model \tabularnewline

$\observe$ & Parameters governing the distribution of photometric uncertainties \tabularnewline \tabularnewline

& \emph{Latent parameters} \tabularnewline

$\sps_{1:N}$ & Stellar population parameters describing rest-frame spectrum (per galaxy) \tabularnewline

$z_{1:N}$ & Redshift (per galaxy) \tabularnewline \tabularnewline

& \emph{Derived quantities} \tabularnewline

$\bar{N}=\bar{N}(\Phi_0, \hyper, \nuisance, \observe)$ & Expected number of selected galaxies given the population, selection and data models \tabularnewline

$\mathbf{f}_{1:N} = \mathbf{f}(\sps_{1:N}, z_{1:N})$ & Model fluxes determined by SPS model (per galaxy) \tabularnewline \tabularnewline

& \emph{Data} \tabularnewline

$\data_{1:N}$ & Data-vector of measured fluxes (per galaxy)\tabularnewline

$\noise_{1:N}$ & Flux measurement uncertainties (per galaxy)\tabularnewline

$N$ & Observed number of selected galaxies \tabularnewline \tabularnewline

& \emph{Selection} \tabularnewline

$S_{1:N}$ & Selection into the sample based on photometric cuts (per galaxy) \tabularnewline

$S^{(k)}_{1:N}$ & Color-based selection into tomographic bin $k$ (per galaxy) \tabularnewline \tabularnewline

\hline
\end{tabularx}}
\caption{Notation for all model parameters included in the forward model.}
\label{tab:parameters}
\end{table*}
\subsection{Generative model}
\label{sec:generate}
The forward model proceeds as follows:
\begin{enumerate}
    \item Draw SPS parameters $\sps$ and redshifts $z$ from the \emph{population model} $P(\sps, z | \hyper)$;
    \item Compute model photometry given the SPS parameters, using the SPS model, $\flux \equiv \flux_\mathrm{SPS}(\sps, z)$;
    \item Draw uncertainties from the \emph{uncertainty model} $P(\noise | \flux, \observe)$;
    \item Draw noisy (calibrated) photometry $\data$ given the model fluxes and uncertainties, from the \emph{data-model} $P(\mathbf{d} | \flux(\sps, z), \noise, \eta)$;
    \item Apply selection cuts on the noisy photometry, where $P(S | \data, \noise)$ equals one (zero) for passing (failing) selection;
    \item Assign tomographic bin label based on the photo-$z$ estimator, $\hat{z}(\data, \noise)$.
\end{enumerate}
The result is a catalog of noisy photometry for selected galaxies, with associated tomographic bin labels.

Repeating the above process until one obtains the same number $N$ of selected objects as in the observed catalog provides a draw from the assumed generative model for the data conditioned on $N$ selected objects\footnote{Note this is implicitly marginalized over the expected present-day volume number density of galaxies; see Appendix \ref{sec:posterior_derivation}}. This can hence be used as a generative model for inferring the hyper-parameters via simulation-based inference, or as the basis for Bayesian hierarchical inference, as described below in \S \ref{sec:inference}.

Repeating this process in the limit of $N\rightarrow\infty$ selected samples, and examining the redshift distributions of the galaxies that make it into each bin, provides the tomographic redshift distributions implied for a given set of modeling and hyper-parameter assumptions. Note that the target redshift distributions are given by a (typically intractable) integral over the population, data, and selection models. The tomographic redshift distribution $n_k(z)$ for galaxies passing selection both into the analysis sample, and subsequently into the $k$-th tomographic bin, is given by:
\begin{align}
\label{nz}
    n_k(z) & \equiv P(z | S^{(k)}, \hyper, \nuisance, \observe) \nonumber \\
    & = \int P(\sps, z | \hyper, S^{(k)}, \noise, \nuisance) \nonumber \\
    &\phantom{= = P(\sps, z | \hyper)}\times P(\noise | S^{(k)}, \hyper, \nuisance, \observe) d\sps\, d\noise \nonumber \\
    & = \int \frac{P(\sps, z | \hyper) P(S^{(k)} | \sps, z, \noise, \nuisance)}{P(S^{(k)} | \hyper, \noise, \nuisance)} \nonumber \\
    &\phantom{= = P(\sps, z | \hyper)}\times P(\noise | S^{(k)}, \hyper, \nuisance, \observe) d\sps\, d\noise \nonumber \\
    & = \int \frac{P(\sps, z | \hyper) P(S^{(k)} | \data, \noise)}{P(S^{(k)} | \hyper, \noise, \nuisance)}P(\data|\sps, z, \noise, \nuisance) \nonumber \\
    &\phantom{= = P(\sps, z | \hyper)}\times P(\noise | S^{(k)}, \hyper, \nuisance, \observe) d\data\,d\sps\, d\noise,
\end{align}
where $S^{(k)}$ denotes selection into both the analysis sample and $k$-th tomographic bin. Forward simulating from the generative model described above provides a way of estimating the target redshift distributions for a given set of hyper-parameters, without the need for direct integration.
\subsection{Emulation of SPS models}
Any applications of this forward model -- either simulation of large mocks, or MCMC sampling the associated posterior -- will require a vast number of SPS model calls. This is only made tractable by neural emulation of SPS models \citep{alsing2020}, which speeds-up SPS computations by a factor of $10^4$ compared to FSPS \citep{conroy2010fsps}.
\subsection{Inference}
\label{sec:inference}
Inference of redshift distributions under the generative model described in \S \ref{sec:generate} requires inferring the population- and data-model parameters $\hyper$ and $\nuisance$ under the forward model assumptions, which in turn provide marginal posteriors for the tomographic redshift distributions via Equation \eqref{nz}\footnote{In practice, the mapping between hyper-parameters and tomographic redshift distributions will be done via simulating large mocks, following \S \ref{sec:generate}}.

In this paper we are focused on validating a baseline forward model for predicting tomographic redshift distributions, without performing inference (or optimization) of the forward model parameters. Nonetheless, it is useful to consider how inference works within our forward modeling framework, and highlight the advantages of performing redshift distribution inference in this fashion.

The joint posterior for the generative model described in \S \ref{sec:generate} is given by (see appendix \ref{sec:posterior_derivation} for a derivation):
\begin{align}
\label{joint_posterior}
P(\hyper, \nuisance, & \{\sps, z\}_{1:N} | \{\data, \noise, S\}_{1:N}, N, \mathcal{O}) = \nonumber \\ 
&P(\hyper) P(\nuisance)\times \prod_{i=1}^{N}\frac{P(\sps_i, z_i | \hyper)P(\data_i | \sps_i, z_i, \noise_i, \nuisance)}{P(S_i | \hyper, \noise_i, \nuisance)}.
\end{align}
where the selection term in the denominator is given by:
\begin{align}
\label{pdet}
    P(S | \hyper, \noise, \nuisance) = \int &P(S | \data, \noise) P(\data | \sps, z, \noise, \nuisance) \nonumber \\
    &\times P(\sps, z | \hyper)\,d\data \,d\sps \,dz.
\end{align}
This joint posterior can then be sampled (using MCMC) to jointly infer the population- and data-model parameters, and SPS parameters and redshifts for each galaxy. The marginal posterior over the hyper- and data-model parameters then provides a posterior over the target tomographic redshift distributions, via Equation \eqref{nz}. 

Phrasing the redshift distribution inference task as a hierarchical model in this way has a number of advantages. Firstly, note that so far this model contains only photometry: the method does not explicitly require spectroscopic calibration data. Where spec-$z$s (or other spectroscopically-derived constraints on SPS parameters) are available for some sub-sets of the galaxies, they can be straightforwardly included by simply appending additional (sharply-peaked) likelihoods for those galaxies. Importantly, the fact that any external spec-$z$ calibration data are not representative of the main sample is unimportant in this approach, provided selection cuts are only performed with respect to the main survey data.

Similarly, inclusion of additional data (e.g., additional bands) from external surveys for subsets of galaxies is also straightforwardly achieved by simply appending additional likelihood-terms for those galaxies. Again, the fact that auxiliary data is only available for biased (unrepresentative) sub-sets of the galaxies is not important, provided selection is not performed with respect to those auxiliary data.

Note that the population model $P(\sps, z | \hyper)$ that appears in Equation \eqref{joint_posterior} is a ``global" quantity: it describes the statistical properties of the background galaxy population (without selection effects), and is therefore the same for all tomographic bins and across all surveys, regardless of their differing selection functions. This opens up strong synergies with the galaxy evolution community who are concerned with constraining (various aspects of) $P(\sps, z | \hyper)$: as our understanding of the statistical properties of the galaxy population improves, this can be fed directly into improved photometric redshift inferences via improved priors on the hyper-parameters $\hyper$.

Finally, since population- and data-model parameters are inferred from the photometric data in a self-consistent fashion, uncertainties in the population and data model parameters will be fully propagated through to the final $n(z)$ inferences.

However, MCMC sampling the joint posterior in Equation \eqref{joint_posterior} requires computing the selection integral in Equation \eqref{pdet} for every galaxy in the sample, in every likelihood evaluation. This presents a severe computational bottleneck for sampling-based methods. In practise, for sampling to be computationally tractable the selection integral will require replacement with a fast emulator (e.g., \citealp{talbot2022}).

Alternatively, simulation-based inference (SBI, aka likelihood-free inference) provides a framework for performing Bayesian inference under complex forward models using only simulations, bypassing the need to compute the likelihood (and selection integral) entirely (e.g., \citealp{alsing2018, alsing2019, jeffrey2020}). For a recent application of SBI to a population model with selection effects, see \citet{gerardi2021}.

\section{Baseline forward model}
\label{sec:forward_model}
In this paper we are focused on demonstrating the ability of a baseline forward model to recover redshift distributions, without performing additional inference or optimization of population-level parameters. We lay out the baseline forward model assumptions for the SPS model in \S \ref{sec:sps}, the galaxy population model in \S \ref{sec:population}, and the data model in \S \ref{sec:data-model}.
\subsection{Stellar population synthesis (SPS) model}
\label{sec:sps}
\begin{table*}
\centering
\scalebox{0.95}{
\begin{tabularx}{\textwidth}{ccc}
\toprule
Parameter & Description & Limits \tabularnewline
\hline \tabularnewline

$\log10 (M/M_\odot)$ & Stellar mass & $\left[7, 13\right]$ \tabularnewline

$\log10 Z_\mathrm{gas}$ & Gas phase metallicity & $\left[-1.98, 0.5\right]$ \tabularnewline

$\log10 u$ & Gas ionization parameter & $\left[-4, -1\right]$ \tabularnewline

$\tau_1$ & Birth cloud (stars younger than $10\mathrm{Myr}$) dust attenuation & $\left[0, 2\right]$ \tabularnewline

$\tau_2$ & Diffuse dust attenuation & $\left[0, 2\right]$ \tabularnewline

$\delta$ & Negative offset from the Calzetti dust attenuation index & $\left[0, 0.4\right]$ \tabularnewline

$\alpha,\,\beta$ & Indices of double-power law star-formation history & $\left[10^{-3}, 10^1\right]$ \tabularnewline

$\tau$ & Transition-time of the double power-law star formation history, as fraction of lookback-time & $\left[0.007, 1\right]$ \tabularnewline

$z$ & Redshift & $\left[0, 2.5\right]$ \tabularnewline \tabularnewline

\hline
\end{tabularx}}
\caption{SPS model parameters and their prior ranges.}
\label{tab:sps_parameters}
\end{table*}
We assume an SPS model with nine free parameters, summarized in Table \ref{tab:sps_parameters} and described below.

Star formation histories (star formation rates as a function of time) are parameterized by a double power law:
\begin{align}
\dot{M}(t ; \alpha, \beta, \tau, z) \propto \frac{M}{\left(t/t^*\right)^\alpha + \left(t/t^*\right)^{-\beta} },
\end{align}
where the transition time is defined as $t^* \equiv \tau t_\mathrm{univ}(z)$ for lookback-time $t_\mathrm{univ}(z)$, and the SFH is normalized such that it integrates to give the total stellar mass $\int_0^{t_\mathrm{univ}(z)}\dot{M}dt=M$. We define star formation rate (SFR) as the average of $\dot{M}$ over the past $100\mathrm{Myr}$.

The gas-phase metallicity $\log10\,Z_\mathrm{gas}$ is a free parameter, and the gas ionization parameter $u$ is set so that it tracks the star-formation rate, assuming the \cite{kaasinen18} relation between gas-ionization and SFR.

The metallicity history of the stellar population assumed to build up with stellar mass production, such that present-day stellar and gas-phase metallicities are identical:
\begin{align}
    Z(t) = (Z_\mathrm{gas} - Z_\mathrm{min})\frac{1}{M}\int_0^t \dot{M}(t)dt + Z_\mathrm{min},
\end{align}
where $Z_\mathrm{min}$ is the minimum metallicity covered by the stellar templates.

Dust attenuation is modelled with two components describing birth cloud (stars younger than ten million years) and diffuse dust screens respectively, following \citet{charlot2000} (see \citealp{leja2017} for details). The birth cloud ($\tau_1$) and diffuse ($\tau_2$) attenuation, and the power law index $n$ of the \citet{calzetti2000} attenuation curve for the diffuse component, are all free model parameters.

We assume MIST stellar evolution tracks and isochrones \citep{choi2016, dotter2016} (based on MESA; \citealp{paxton2010, paxton2013, paxton2015}). Nebular line and continuum emission is generated with CLOUDY \citep{ferland2013}, using model grids from \citet{byler2017}.

We emulate the photometry (apparent magnitudes) in all relevant bands using \texttt{speculator} \citep{alsing2020}. The apparent magnitude in each band as a function of the SPS parameters and redshift is parameterized by a dense neural network with four hidden layers of 128 units each, and activation functions described in \citep{alsing2020}. Each emulator is trained on $6.4\times 10^6$ training samples, with SPS parameters and redshifts drawn from the population model (see \S \ref{sec:population} below) and model photometry computed using FSPS \citep{conroy2010fsps}. Training is performed following the prescription in \citep{alsing2020}, and we ensure that the $99.9\%$ intervals for the emulator error distributions are better than $2\%$ in all bands.
\subsection{Galaxy population model}
\label{sec:population}
Specifying a population model for the SPS parameters and redshift amounts to specifying the joint (prior) distribution that characterizes the statistical properties of the galaxy population. Galaxy formation and evolution physics results in complex relationships between the stellar population parameters. In an effort to capture as much of this phenomenology as possible, we factorize the population prior into the following (generic) structure:
\begin{align}
    P(\mathrm{SPS\;parameters},&\;\mathrm{redshift}) = \nonumber \\
    & P(\mathrm{mass,\;redshift}) \nonumber \\
    & P(\mathrm{metallicity} | \mathrm{SFR}, \mathrm{mass}) \nonumber \\
    & P(\mathrm{SFR} | \mathrm{mass}, \mathrm{redshift}) \nonumber \\
    & P(\mathrm{dust} | \mathrm{SFR}, \mathrm{mass}, \mathrm{metallicity}) \nonumber \\
    & P(\mathrm{age} | \mathrm{SFR}, \mathrm{mass}).
\end{align}
Factorized this way, the population model is decomposed into a number of well-studied relations between galaxy characteristics: $P(\mathrm{mass,\;redshift})$ is given by the redshift-evolving mass-function, $P(\mathrm{metallicity} | \mathrm{SFR}, \mathrm{mass})$ characterizes the fundamental metallicity relation, $P(\mathrm{SFR} | \mathrm{mass}, \mathrm{redshift})$ characterizes the star-forming sequence, $P(\mathrm{dust} | \mathrm{SFR}, \mathrm{mass}, \mathrm{metallicity})$ specifies the relationship between dust and the star-formation and chemical enrichment histories, and $P(\mathrm{age} | \mathrm{SFR}, \mathrm{mass})$ the empirical relationship between ages and star-formation histories.

For the SPS model set-up chosen for this study (summarized in Table \ref{tab:sps_parameters}), the specific population model assumptions and default parameters are taken as follows.
\subsubsection{Mass function}
\begin{figure*}
\centering
\includegraphics[width = 17.5cm]{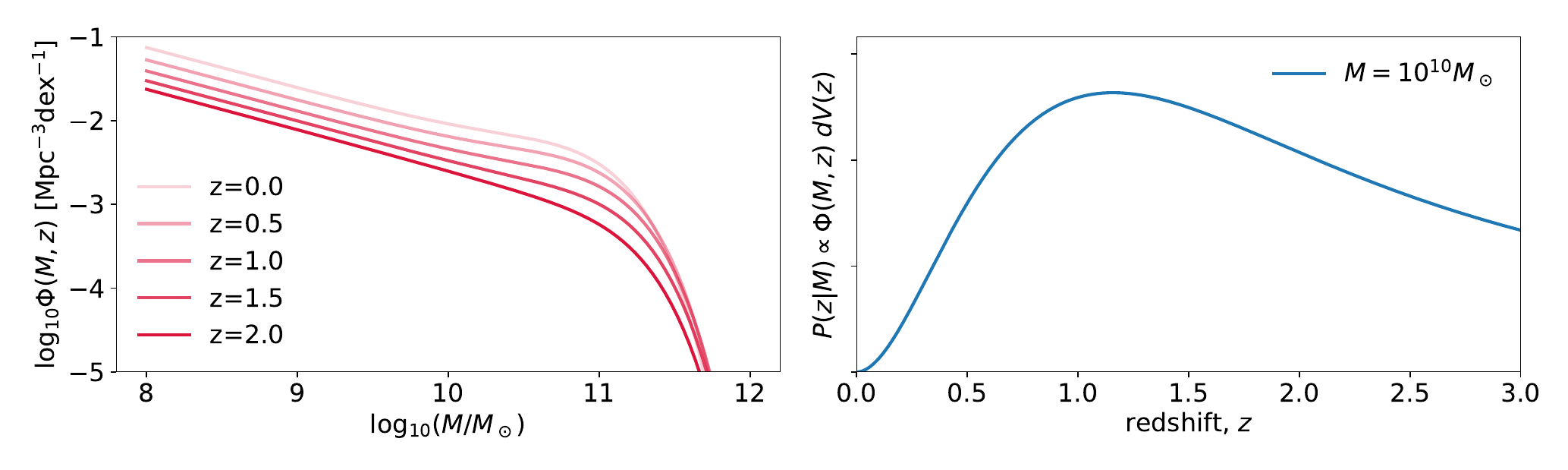}
\caption{Left: Redshift-evolving mass function from \citet{leja2020MF}. Right: Implied redshift distribution for the background galaxy population (without selection) at $M=10^{10}M_\odot$.}
\label{fig:mf}
\end{figure*}
The joint distribution of mass and redshift is defined by:
\begin{align}
    P(M, z) \propto \Phi(M, z) dV(z),
\end{align}
where $\Phi(M, z)$ is the un-normalized mass function, and $dV(z)$ the differential comoving volume element. For the mass-function, we assume a mixture of two Schechter functions, with default parameter values and redshift evolution taken from \citet{leja2020MF}. We assume a \emph{Planck} 2015 \citep{planck2015} cosmology for the comoving volume element. The assumed mass-function and redshift prior is show in Figure \ref{fig:mf}.

\subsubsection{Fundamental metallicity relation (FMR)}
Galaxies undergo continuous chemical evolution, as heavier elements are produced in stars and expelled into the interstellar medium, and gas flows regulate the metal content by either dilution or expulsion of enriched gas out of the galactic potential. On global scales, this results in a tight relationship between the gas-phase metallicity and star formation history of a galaxy (e.g., mass and SFR) -- the so-called fundamental metallicity relation \citep{yates2012, andrews2013, nakajima2014, yabe2015, salim2014, salim2015, kashino2016, cresci2019, curti2020}. Qualitatively, galaxies on the FMR tend towards lower metallicities for higher star formation rates, and higher metallicities for higher masses, but the overall shape is a non-linear function of both mass and SFR. The FMR is typically considered to be independent of redshift, with galaxies moving along the relation as they evolve (and preferentially occupying different regions of the relation at different redshifts), but the relation itself being constant over cosmic history\footnote{The physics governing chemical enrichment is assumed to be constant over cosmic time.}.

\begin{figure}
\centering
\includegraphics[width = 8cm]{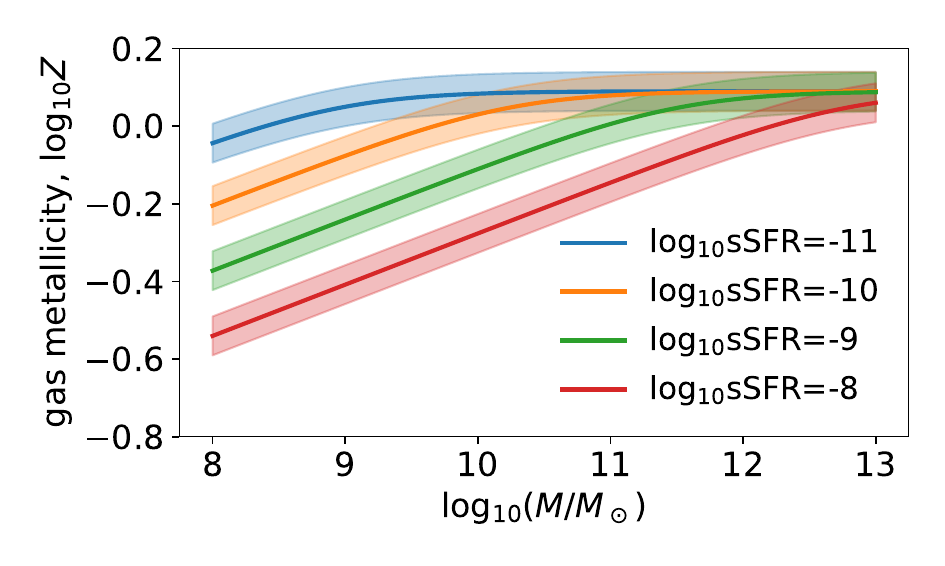}
\caption{Fundamental metallicity relation prior on gas-phase metallicity conditioned on mass and specific star formation rate (in units of $\mathrm{Gyr}^{-1}$). Solid lines represent the mean, and the bands show the FWHM of the assumed student-t distribution.}
\label{fig:fmr}
\end{figure}

We take the FMR parameterization and default parameter values from \citet{curti2020}, where the FMR was measured over the broad stellar-mass and SFR ranges covered by SDSS. The median gas-phase metallicity as a function of mass and SFR is parameterized as:
\begin{align}
\label{fmr}
    &\langle\log10 Z_\mathrm{gas}\rangle = \Delta Z_0 - \frac{\gamma }{\beta} \log10\left[1 + \left(\frac{M}{M_0(\mathrm{SFR})}\right)^{-\beta}\right],
\end{align}
where
\begin{align}
    &\log10 M_0(\mathrm{SFR}) = m_0 + m_1\log10(\mathrm{SFR}),
\end{align}
with default parameters $\Delta Z_0 = 0.09$, $\gamma = 0.3$, $m_0 = 10.1$, $m_1 = 0.56$, and $\beta = 2.1$. We assume a Student's-t distribution\footnote{Hereafter, student-t.} (with two degrees-of-freedom) for $P(\log10 Z_\mathrm{gas} | M, \mathrm{SFR})$, where the mean is given by Equation \eqref{fmr}, and the FWHM is equal to $0.05$. The FMR is shown in Figure \ref{fig:fmr}.

Measurement of the FMR is sensitive to the way in which SFRs and metallicities are estimated (see e.g., \citealp{telford2016} and \citealp{ cresci2019}), with the SFRs and metallicities used to calibrate the FMR from spectra and those arising in a given SPS model being subtly different proxies for those quantities. Therefore, when fitting the FMR to photometric data as part of the population model, it would be prudent to set reasonably broad priors on the FMR parameters in order to capture any biases relative to measurements based on spectra.

While the majority of galaxies are expected to live on the FMR (since the same physics drives chemical enrichment for most galaxies), those processes will be disrupted in merger events. Merged galaxies are therefore not expected to follow the same FMR relation \citep{bustamante2020}. This can be compensated for by adding heavy tails to the FMR (as we do here with the student-t distribution), or deriving a separate model for the metallicities of merged galaxies.
\subsubsection{Star forming sequence}
\label{sec:sfs}
The star-forming sequence (SFS) characterizes the (redshift-evolving) relationship between star formation rate and mass, with the vast majority of galaxies forming most of their mass either on or passing through the star-forming sequence \citep{leitner2012, abramson2015}. Qualitatively, the SFS is characterized by star-forming and quiescent galaxies with ongoing and negligible star formation rates respectively. The clustering of galaxies into those two populations with different characteristic SFRs results in a highly non-Gaussian -- sometimes bimodal -- distribution of SFRs (conditioned on mass and redshift) in the galaxy population as a whole \citep{daddi2007, noeske2007, karim2011, rodighiero2011, whitaker2012, whitaker2014, speagle2014, renzini2015, schreiber2015, tomczak2016, leslie2020, leja2021SFS}.

In this work, we base our SFS model on the measured relation from \citet{leja2021SFS}, who fit a normalizing flow to learn $P(\mathrm{SFR} | M, z)$ from 3D HST data. The galaxy sample used in \citet{leja2021SFS} is mass-complete down to around $10^9 M_\odot$ over the redshift range relevant for this study ($z\leq 1.5$). We note that $P(\mathrm{SFR} | M, z)$ is a very smooth function of mass and redshift, and to ensure sensible extrapolation below the mass limit we fit a surrogate model (with good extrapolation properties) to the normalizing flow of \citet{leja2021SFS} (see Figure \ref{fig:sfs}, and Appendix \ref{sec:sfs_surrogate} for technical details).

\begin{figure}
\centering
\includegraphics[width = 8cm]{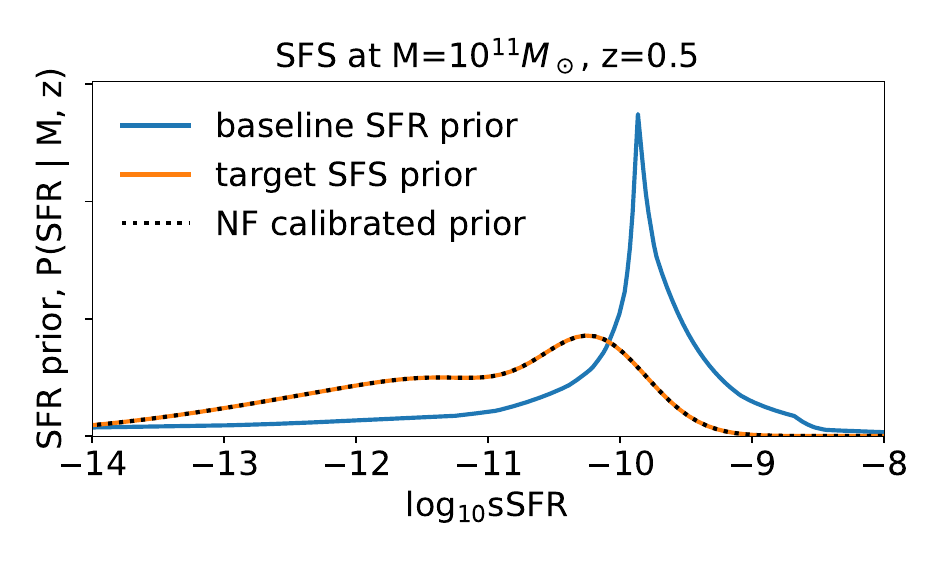}
\caption{The prior on the star formation rate implied by taking uniform priors over the double power-law SFH parameters $(\log10 \alpha, \log10 \beta, \tau)$ over the ranges specified in Table \ref{tab:sps_parameters} is shown in blue, while the target prior (specified by the SFS measurement of \citet{leja2021SFS}) is shown in orange. The SFR prior obtained by re-calibrating the baseline prior with a normalizing flow (as described in \ref{sec:sfs}) is shown by the black dotted line, giving excellent agreement with the target SFS prior.}
\label{fig:sfs_calibrated}
\end{figure}

\begin{figure*}
\centering
\includegraphics[width = 17.5cm]{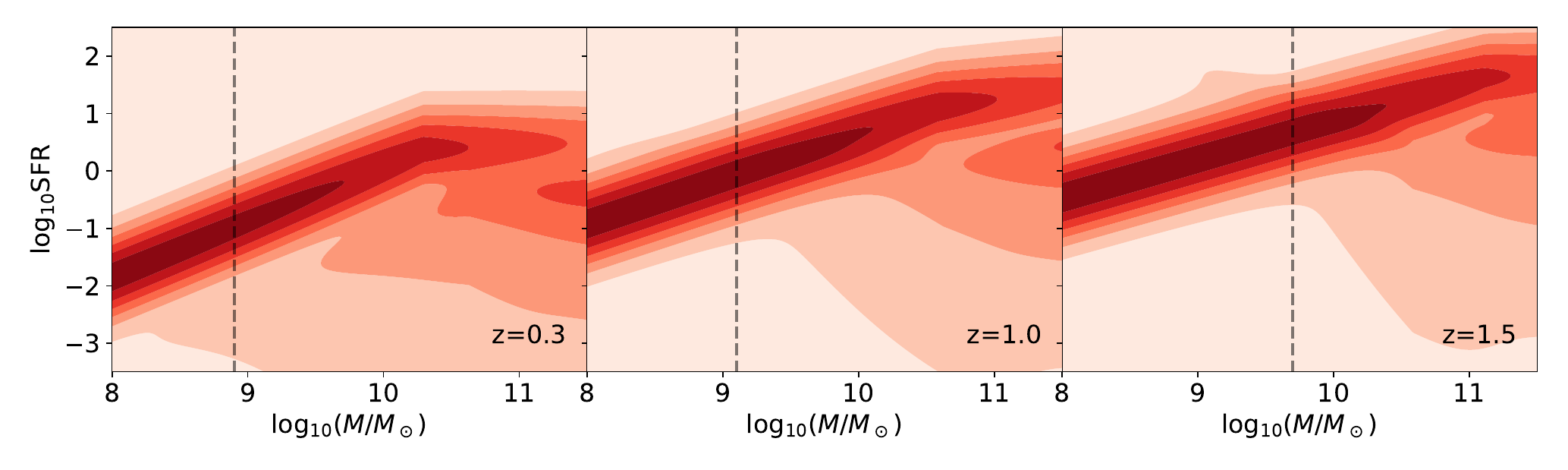}
\caption{Star forming sequence prior on star formation rate (in units of $M_\odot\mathrm{yr}^{-1}$) conditioned on mass and redshift, based on the measurement from \citet{leja2021SFS} (see Appendix \ref{sec:sfs_surrogate} for details). The vertical lines represent the mass-complete limit for the 3D HST data on which the \citet{leja2021SFS} fit was performed; we extrapolate the SFS below the mass limit as shown in the Figure.}
\label{fig:sfs}
\end{figure*}

Specifying a population model requires that we put a prior on the SFH parameters (in this case, the three parameters of the double-power SFH), but the star-forming sequence provides only a prior constraint on a derived quantity: the star formation rate, $\mathrm{SFR}(\alpha, \beta, \tau, z)$. Hence, we need to define our prior over the SFH parameters in such a way that the target prior over the SFR is satisfied. One of the problems with parametric SFH models such as the double power-law is that simple priors on the SFH parameters lead to strong (and undesirable) implied priors on derived quantities such as SFR (e.g., \citealp{carnall2019}). For example, taking a baseline uniform prior in $(\log10 \alpha, \log10 \beta, \tau)$ over reasonable ranges (c.f. Table \ref{tab:sps_parameters}) leads to the undesirable implicit SFR prior $P_0(\mathrm{SFR} | z)$ shown in Figure \ref{fig:sfs_calibrated}.

To ensure that our SFH priors \emph{only} encode the SFS, without a spurious additional contribution from any baseline prior assumptions, we define the SFH priors as:
\begin{align}
    P(\alpha, \beta, \tau | M, z) = \frac{\pi_0(\alpha, \beta, \tau)\;P(\mathrm{SFR} | M, z)}{P_0(\mathrm{SFR} | z)}
\end{align}
where $\pi_0(\alpha, \beta, \tau)$ is the baseline (uniform) prior on the SFH parameters, $P(\mathrm{SFR} | M, z)$ is the target SFS prior, and $P_0(\mathrm{SFR} | z)$ is the implicit prior on SFR implied by $\pi_0$. The implicit SFR prior is defined by the surface integral
\begin{align}
\label{surface}
P_0(\mathrm{SFR} | z) = \int_{\mathrm{SFR}=\mathrm{const.}} \pi_0(\alpha, \beta, \tau) dS,
\end{align}
where $dS$ is the surface element in the SFH parameter space $(\alpha, \beta, \tau)$.

Dividing out the implicit SFR prior in this way insures that the overall prior on SFR is specified by the SFS only. To avoid having to compute the surface integrals in Equation \eqref{surface} directly, we train a normalizing flow to learn the conditional density $P_0(\mathrm{SFR} | z)$ so that it can be conveniently divided out (technical details are given in Appendix \ref{sec:nf_priors}).
\\
\\
\subsubsection{Dust}
\begin{figure}
\centering
\includegraphics[width = 8cm]{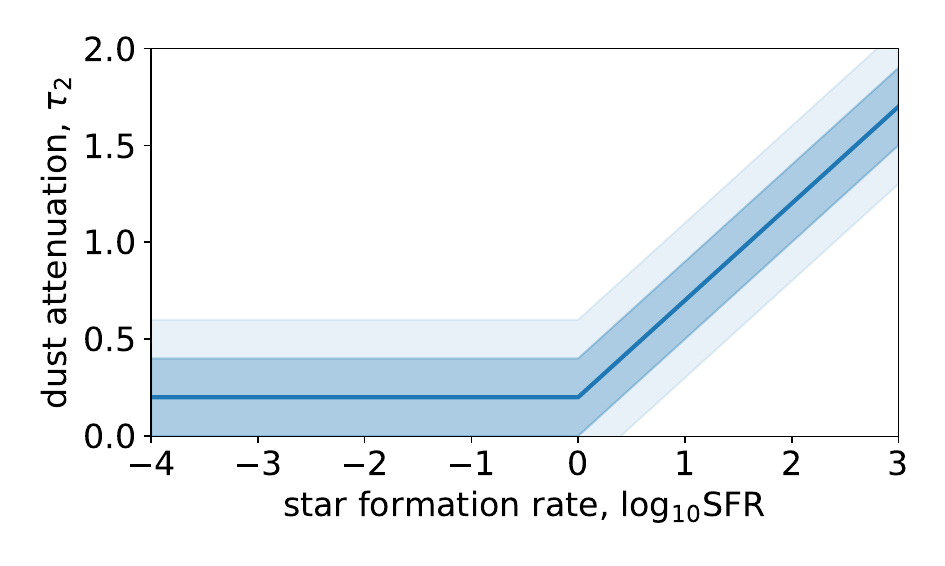}
\caption{Prior on the diffuse dust attenuation conditioned on SFR. Solid line shows the mean, and the bands show the 1- and 2-$\sigma$ contours.}
\label{fig:dust}
\end{figure}
The amount of dust attenuation, and the shape of the effective attenuation law, is governed by the total amount of dust, the grain composition, the dust-star-gas geometry in the galaxy, and its inclination relative to the observer. This can be encoded as a relationship between dust attenuation parameters and star formation histories (SFR and mass), as well as potentially metallicity and redshift (see \citealp{salim2020} for a review).

We take a relatively simple dust prior model, where the dust attenuation in the diffuse component is assumed to scale with SFR according to (following \citealp{tanaka2015}):
\begin{align}
    \langle \tau_2 \rangle = 0.2 + 0.5\,\log10\mathrm{SFR}\; \Theta(\log10\mathrm{SFR}),
\end{align}
where $\Theta$ is the Heaviside step function. We assume a Gaussian prior on the diffuse dust attenuation with the mean given above, and standard deviation of $0.2$. The dust attenuation prior is shown in Figure \ref{fig:dust}.

The index of the dust attenuation law (for the diffuse component) is assumed to vary as a function of the total dust attenuation, with mean given by:
\begin{align}
    \langle \delta \rangle = -0.095 + 0.111 \tau_2- 0.0066 \tau_2^2,
\end{align}
where $\delta$ is the (negative) offset from the index of the Calzetti attenuation curve \citep{calzetti2000}. We take a Gaussian prior on $\delta$, with mean given above and standard deviation $0.4$.

For the birth cloud component, we take a Gaussian prior on the ratio $\tau_1 / \tau_2$ of the birth cloud and diffuse dust components with mean equal to $1$ and standard deviation $0.3$. This is consistent with previous findings that the dust optical depth in nebular emission lines is roughly twice that of the stellar component \citep{calzetti1994, price2014, reddy2015}.

We leave more sophisticated dust prior modeling, e.g., where dust attenuation properties are conditioned on SFR, mass, metallicity and redshift (e.g., \citealp{nagaraj2022}), to future work.
\subsubsection{Age}
The double power-law SFH parameterization and prior implicitly links age and SFR, with older (younger) galaxies having lower (higher) star-formation rates, qualitatively in line with expectations. However, the assumed priors on the double-power-law SFH parameters allow for a tail down to very low ages; we therefore impose a lower cut of one $\mathrm{Gyr}$ on the galaxy ages\footnote{We take age here to mean mass-weighted age.} to eliminate spuriously young galaxies. We do not impose any additional priors on age.

\subsection{Data model}
\label{sec:data-model}
The data-model characterizes the sampling distribution of the measured photometry, given the true (model) fluxes, encoding both calibration (i.e., zero points), modeling errors, and measurement noise. We treat the sampling distribution of observed fluxes as a student-t distribution with two degrees-of-freedom,
\begin{align}
    P(\data | \flux, \noise, \nuisance) \propto \prod_{b=1}^{N_\mathrm{bands}} \left[1+\frac{1}{2}\left(\frac{d_b - \alpha_{\mathrm{zp},b} f_b}{\Sigma_b(f_b)}\right)^2\right]^{-3/2},
\end{align}
where the total uncertainty $\Noise$ is given by:
\begin{align}
    \Sigma_b^2 = \sigma_b^2 + (\beta_b\alpha_{\mathrm{zp},b}f_b)^2.
\end{align}
The data-model parameters $\nuisance = (\boldsymbol\alpha_\mathrm{zp},\,\boldsymbol\beta)$ characterize the zero points $\boldsymbol\alpha_\mathrm{zp}$, and error floors $\boldsymbol\beta$ (encoding modeling errors, emulator errors, and an effective noise floor on the measurement uncertainties).

In the application to GAMA and VVDS data in \S \ref{sec:GAMA}-\ref{sec:VVDS}, we fix the zero points to the values published by the respective survey collaborations, and assume a default value of $0.03$ for the (fractional flux) error floors in all bands.

We note that this data model is readily extendable to include additional error terms for emission line modeling errors, SED modeling-error as a function of rest-frame wavelength, and parameters describing the shape (e.g., skewness or tailweight) of the data sampling distribution. A more sophisticated error-model (including hierarchical calibration of hyper-parameters) is investigated in our companion paper \citet{leistedt2022}.

\subsection{Uncertainty model}
\label{sec:uncertainty-model}
The uncertainty model describes the distribution of photometric measurement uncertainties over the survey, which will vary from galaxy to galaxy due to heterogeneous observing conditions and strategy, varying difficulty in extracting photometry from galaxies images with different morphologies and geometries, and will also scale with the (true) flux owing to the Poisson photon count contribution to the overall measurement error.

We take a data-driven approach to uncertainty modeling, where we learn the distribution $P(\noise | \flux, \observe)$ directly from the data. This process proceeds in two steps. First, we fit each of the galaxies under the SPS model, population prior and data model assumptions described above, by MCMC sampling their individual posteriors (c.f. Equation \eqref{joint_posterior} with fixed hyper-parameters). For each galaxy, this provides a maximum a posteriori (MAP) estimate for their true fluxes $\flux$. We then take the catalog of MAP estimated fluxes and associated uncertainties $\{\flux, \noise\}_{1:N}$, and train a Mixture Density Network (MDN) to learn $P(\noise | \flux, \observe)$. The MDN parameterizes the uncertainty distribution as a Gaussian mixture:
\begin{align}
    P(\noise | \flux, \observe) = \sum_{c=1}^{N_\mathrm{comp.}} r_c(\flux;\mathbf{w}) \mathcal{N}\left[\boldsymbol\mu_c(\flux;\mathbf{w}), \boldsymbol\Sigma_c(\flux;\mathbf{w})\right],
\end{align}
where the component weights, means and variances are functions of flux, parameterized by a dense neural network (whose weights and biases are denoted by $\mathbf{w}$). The MDN is trained by minimizing the total (negative) log-likelihood of the data under the model with respect to the network weights\footnote{Note that this is equivalently a Monte Carlo estimate of the Kullback-Leibler divergence between the model and true distribution, up to an additive constant.}:
\begin{align}
    \mathcal{L(\mathbf{w})} = -\sum_{i=1}^{N_\mathrm{galaxies}} \ln\,P(\noise_i | \flux_i, \mathbf{w}).
\end{align}
Throughout this paper we take a default MDN with $12$ components, and a single hidden layer with $256$ units and leaky-ReLU activation functions. Examples of trained MDNs for the uncertainty distributions for GAMA and VVDS are shown in Figures \ref{fig:GAMA_noise} and \ref{fig:VVDS_noise}.
\section{Case study I: GAMA}
\label{sec:GAMA}
The Galaxy And Mass Assembly (GAMA) survey covers $250$ square degrees, and has obtained $\sim 230,000$ spectroscopic redshifts over the past decade \citep{driver2011}. The survey was designed to have simple target selection based on photometry alone (discussed below), making it an ideal dataset for validating our forward model. We take the GAMA data release 4 \citep{driver2022GAMADR4}, with photometry in KiDS $ugri$ and VIKING $ZYHJKs$ bands.

The main selection is performed on the KiDS $r$-band, with spectroscopic redshifts measured for all galaxies with $r < 19.65$. An additional color cut of $(J-Ks) > 0.025$ is made for star-galaxy separation. With these cuts, we are left with a sample of $206,454$ galaxies with 9-band photometry ($ugriZYHJKs$), and measured spectroscopic redshifts (for validation).
\subsection{Data model}
We assume student-t uncertainties on the fluxes as described in \S \ref{sec:data-model}, and take the extinction and zero point corrections provided with GAMA DR4 \citep{driver2022GAMADR4}.

\begin{figure*}
\centering
\includegraphics[width = 17.5cm]{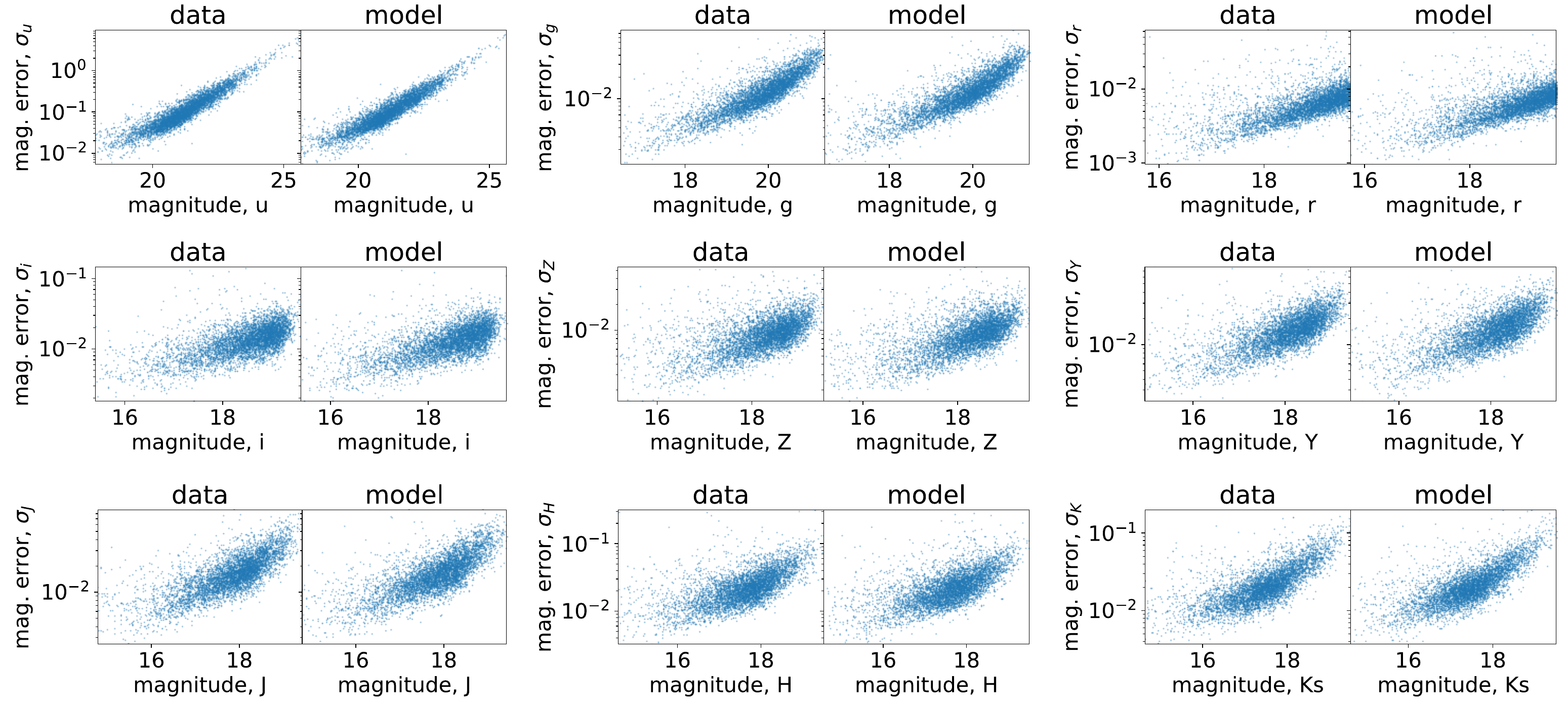}
\caption{Magnitude uncertainties versus magnitudes for the GAMA data (left panels) versus the trained MDN model for the error distribution conditioned on flux (right panels). Note that the magnitudes in both the left and right panels are maximum a posteriori (MAP) magnitudes, from an initial fit of the SPS model to the GAMA galaxies as described in \S \ref{sec:uncertainty-model}.}
\label{fig:GAMA_noise}
\end{figure*}

We train an MDN to model the distribution of measurement uncertainties as a function of flux, trained on the GAMA data, as described in \S \ref{sec:uncertainty-model}. The uncertainty distributions and corresponding trained models are shown side-by-side in Figure \ref{fig:GAMA_noise}.
\subsection{Tomographic binning}
\label{sec:z_estimator}
For the purpose of tomographic binning, we train a simple (dense) neural network estimator for the redshift given the measured KiDS and VIKING photometry. We take a dense network with four hidden layers with $64$ units each and leaky-ReLU activations, passing the measured magnitudes and magnitude errors in the 9-bands as inputs and the estimated redshift as output. The network is trained to minimize the mean square error on the redshift, trained on the GAMA photometry and redshifts. Training is performed with Adam using a batch size of $1024$, a training:validation split of 90:10, and triggering early-stopping when the validation loss has ceased to improve after 30 epochs. The resulting redshift estimator has an overall accuracy of around $\sigma_z\simeq 0.06$.

The GAMA galaxies are binned into two tomographic bins based on their estimated redshift, $0 < \hat{z} < 0.2$ and $\hat{z} > 0.2$ for the two bins respectively.
\subsection{Results}
\label{sec:gama_results}
Forward model predictions are obtained by generating a large mock catalog following the prescription in \S \ref{sec:generate}. We MCMC sample the population model (imposing a prior limit of $r < 20.65$), draw uncertainties and add noise according to the uncertainty and data models described above, and apply selection cuts $r < 19.65$ and $(J-Ks) > 0.025$ to the simulated noisy photometry. Tomographic bin labels are assigned based on the redshift estimator described above. We continue sampling until $5\cdot 10^5$ selected samples are obtained.

\begin{figure*}
\centering
\includegraphics[width = 17.5cm]{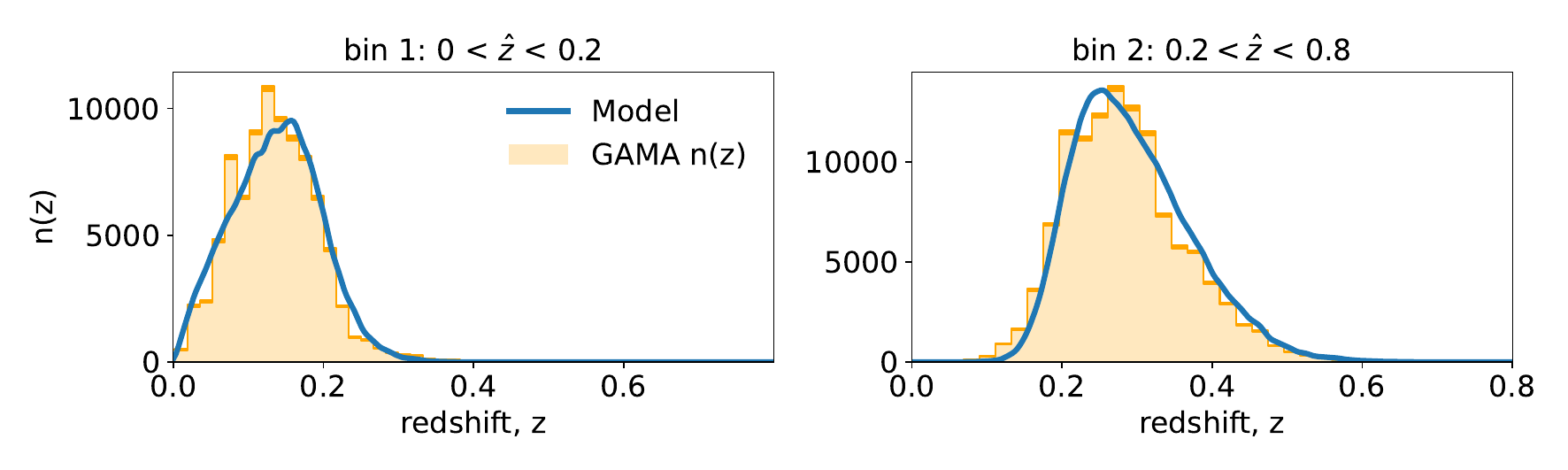}
\caption{Tomographic redshift distributions obtained by the forward model (blue) compared to the histogram of the GAMA spectroscopic redshifts (orange). The model predictions are in excellent agreement with the distributions of the spectroscopic redshifts.}
\label{fig:GAMA_nz}
\end{figure*}

\begin{figure*}
\centering
\includegraphics[width = 17.5cm]{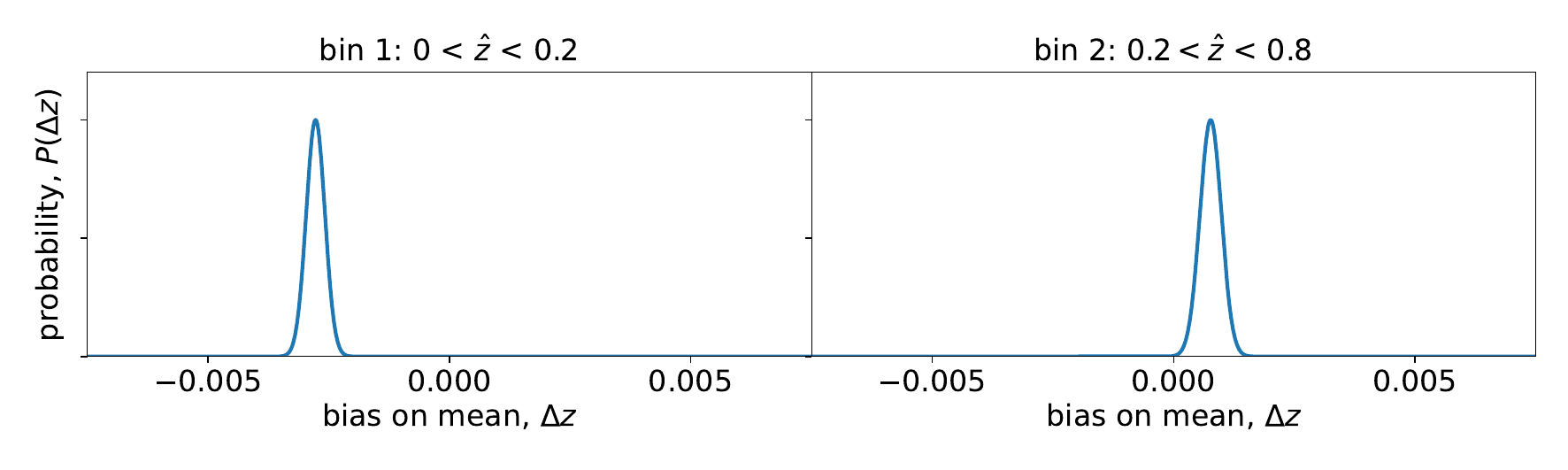}
\caption{Bias on the mean redshift of the model redshift distributions versus the data for GAMA, $\Delta z = \langle z_\mathrm{model}\rangle - \langle z_\mathrm{data}\rangle$. The distributions are obtained by bootstrapping samples from the model $n(z)$, taking a kernel density estimate (KDE) of the bootstrapped sample means, and centering the KDE on the difference between the sample mean of the spec-$z$s and the mean of the model $n(z)$.}
\label{fig:GAMA_deltaz}
\end{figure*}

The tomographic redshift distributions predicted by the forward model are shown alongside the spec-$z$ histograms in Figure \ref{fig:GAMA_nz}, and the corresponding bias on the mean of the redshift distribution in shown in Figure \ref{fig:GAMA_deltaz}. The forward model is able to predict the redshift distributions with a bias of around $0.003$ and $0.001$ on the mean, for the two respective tomographic bins. This is comfortably accurate enough for ongoing Stage III surveys (e.g., \citealp{asgari2021}), where the statistical error on the mean redshift per tomographic bin is $\mathcal{O}(0.01)$, and cosmological parameter constraints should be insensitive to biases of $\lesssim 0.04$ \citep{hildebrandt2016}. The model predictions are very close to the accuracy requirements for Stage IV surveys, where the bias on the mean should not exceed\footnote{For LSST year 1 analysis, the requirement on the mean bias per tomographic bin is $\Delta z < 0.002(1+z)$, decreasing to $\Delta z < 0.001(1+z)$ by year 10 \citep{mandelbaum2018}.} $\Delta z < 0.002(1+z)$ \citep{mandelbaum2018}.

We re-iterate that these model predictions are obtained by assuming (fixed) default parameters for the population model described in \S \ref{sec:population}, fitting the uncertainty distribution to the GAMA data to characterize the distribution of photometric errors (as a function of flux), and assuming the zero-point calibration provided by the GAMA collaboration \citep{driver2022GAMADR4}.

\section{Case study II: VVDS}
\label{sec:VVDS}
Similar to GAMA, the VIMOS VLT Deep Survey (VVDS; \citealp{lefevre2013}) is a spectroscopic survey designed to have simple photometric target selection. We focus on VVDS-\emph{Wide}, which covers 8.7 square degrees and obtained spectra for $25\,805$ galaxies down to $I < 22.5$.

We take photometry in the $BVRI$ bands for VVDS-\emph{Wide}, which were obtained with the CFH12K camera at CFHT. The main selection is performed in the $I$-band, with spectra obtained for objects with $17.5 < I < 22.5$. The imaging survey is sufficiently deep (limiting magnitude of $I = 24.8$) to ensure $100\%$ completeness down to $I = 22.5$ for the spectroscopic sample. Star-galaxy separation was performed on the spectra, so no other photometric cuts were performed. We selected only galaxies with redshift quality flags of 3 or 4 ($>95\%$ probability to obtain a correct redshift according to \citealp{lefevre2013}). The relevant information for modeling any correlations between the assessed reliability and redshift in detail is not publicly available for this catalog, and hence we make no attempt to model this implicit selection effect. However, Fig. 13 in \cite{lefevre2013} for the VVDS-{\it Deep} sample in the same magnitude range suggests that the spectroscopic success rate for this subsample is expected to be roughly uniform up to $z\simeq1$ and drop thereafter. This only impacts $\sim 1-2\%$ of the sample, in a regime where photometric selection is expected to strongly dominate in any case. Hence we do not expect a significant impact on our results from the redshift dependence of the spectroscopic success rate.
\\
\\
\subsection{Data model}
We again assume student-t uncertainties on the fluxes as described in \S \ref{sec:data-model}, and take the extinction and zero point corrections provided by the VVDS team \citep{lefevre2013}.

\begin{figure*}
\centering
\includegraphics[width = 17.5cm]{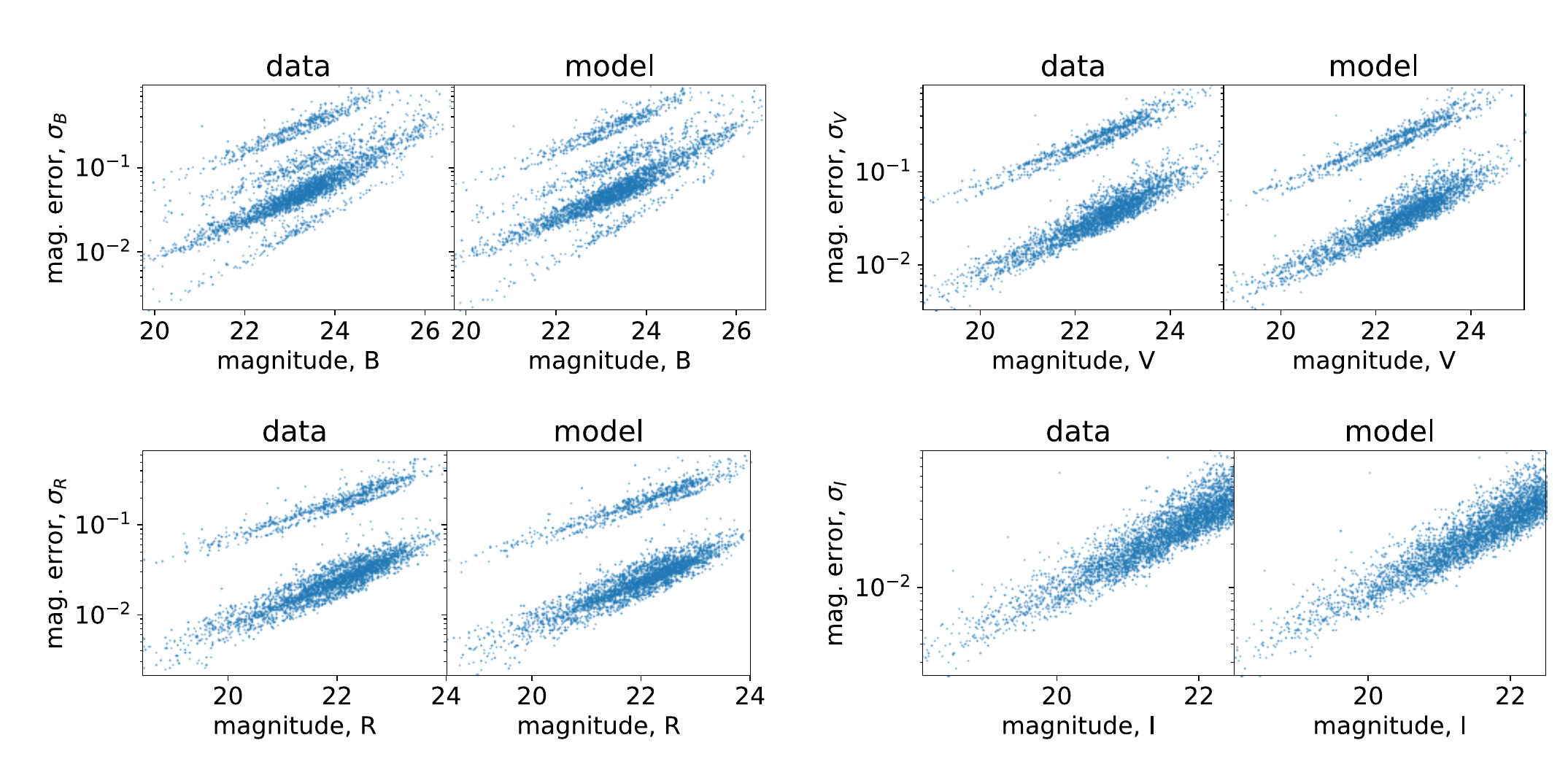}
\caption{Magnitude errors versus magnitudes for the VVDS data (left panels) versus the trained MDN model for the error distribution conditioned on flux (right panels). Note that the magnitudes in both the left and right panels are maximum a posteriori (MAP) magnitudes, from an initial fit of the SPS model to the VVDS galaxies as described in \S \ref{sec:uncertainty-model}.}.
\label{fig:VVDS_noise}
\end{figure*}

We train a MDN to model the distribution of measurement uncertainties as a function of flux, trained on the VVDS data, as described in \S \ref{sec:uncertainty-model}. Unlike GAMA, the VVDS $BVRI$ photometry is patchy with roughly a quarter of objects missing measurements in at least one band. We set the fractional uncertainties for missing values to $1000$, and constrain one component in the mixture model to be a delta function at $1000$, where the relative weight of that component in the MDN then encodes the relative probability of having a missing value, as a function of flux. The uncertainty distributions (for non-missing values) and corresponding trained models are shown side-by-side in Figure \ref{fig:VVDS_noise}. Note the multi-modal structure in the uncertainty distributions, owing to varying depth photometry in all but the $I$-band. This makes for a more challenging test case for the forward modeling framework.
\subsection{Tomographic binning}
For the purpose of tomographic binning, we train a simple (dense) neural network estimator for the redshift given the measured $BVRI$ photometry. We again take a dense network with four hidden layers with $64$ units each and leaky-ReLU activations, passing the measured magnitudes and magnitude errors in the 4-bands as inputs and the estimated redshift as output. The network is trained on the VVDS photometry and spec-$z$s as described in \S \ref{sec:z_estimator}. The resulting redshift estimator has an overall accuracy of around $\sigma_z\simeq 0.2$. Note that the redshift estimator is considerably less accurate in this case comapred to GAMA, owing to the poorer constraining power of the $BVRI$ bands, the prevalence of missing values in the VVDS photometry, and the smaller training set.

The VVDS galaxies are binned into three tomographic bins based on their estimated redshift: $0 < \hat{z} \leq 0.4$; $0.4 < \hat{z} \leq 0.75$; and $0.75 < \hat{z} \leq 2$.
\subsection{Results}
As in \S \ref{sec:gama_results}, model predictions are obtained by MCMC sampling the population model (imposing a prior limit of $I < 23.5$), drawing uncertainties and adding noise according to the uncertainty and data models described above, and applying selection cuts $17.5 < I < 22.5$ to the simulated noisy $I$-band magnitudes. Tomographic bin labels are assigned based on the redshift estimator described above. Sampling is continued until $5\cdot 10^5$ selected samples are obtained.

\begin{figure*}
\centering
\includegraphics[width = 17.5cm]{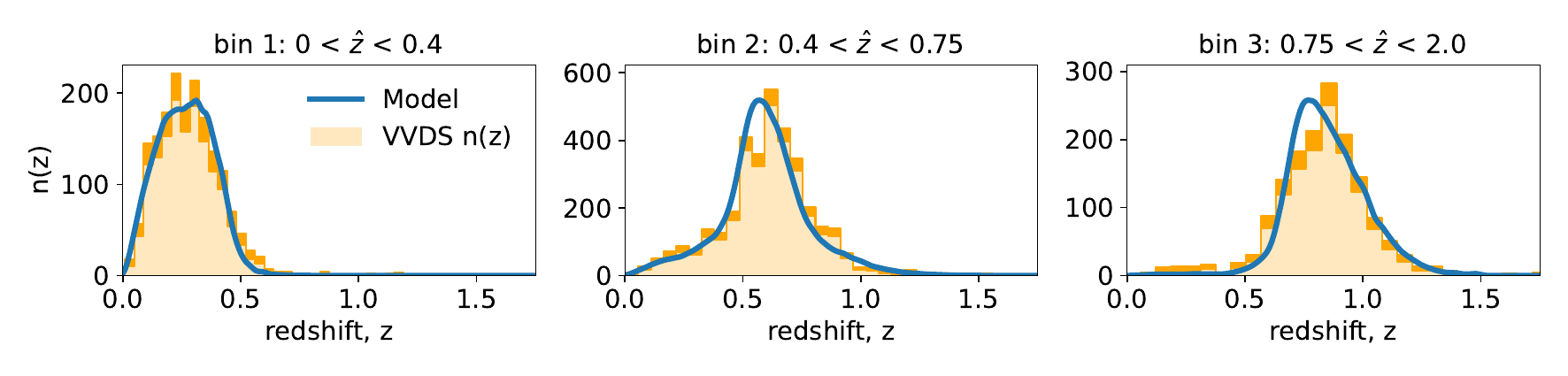}
\caption{Tomographic redshift distributions obtained by the forward model (blue) compared to the histogram of the VVDS spectroscopic redshifts (orange). The width of the histogram bars indicates $\pm \sqrt{N}$ Poisson noise. The model predictions are in excellent agreement with the distributions of the spectroscopic redshifts.}
\label{fig:VVDS_nz}
\end{figure*}

\begin{figure*}
\centering
\includegraphics[width = 17.5cm]{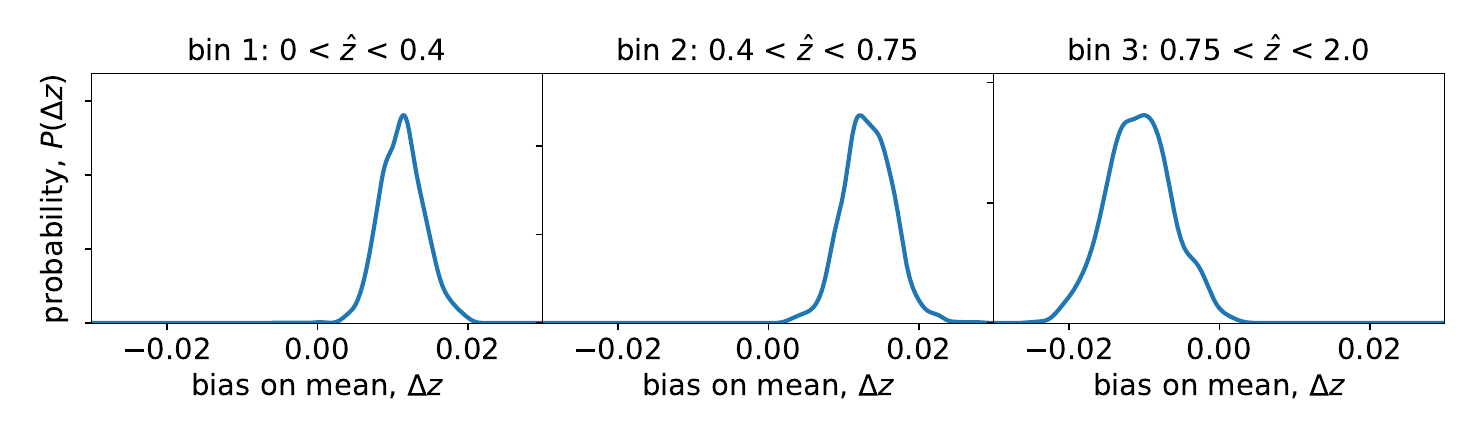}
\caption{Bias on the mean redshift of the model redshift distributions versus the data for VVDS, $\Delta z = \langle z_\mathrm{model}\rangle - \langle z_\mathrm{data}\rangle$. The distributions are obtained by bootstrapping samples from the model $n(z)$, taking a kernel density estimate (KDE) of the bootstrapped sample means, and centering the KDE on the difference between the sample mean of the spec-$z$s and the mean of the model $n(z)$.}
\label{fig:VVDS_deltaz}
\end{figure*}

The tomographic redshift distributions predicted by the forward model are shown alongside the spec-$z$ histograms for VVDS in Figure \ref{fig:VVDS_nz}, and the corresponding bias on the mean of the redshift distribution in shown in Figure \ref{fig:VVDS_deltaz}.

The forward model is able to predict the redshift distributions with a bias of $\Delta z \simeq 0.01$ on the mean in all three bins. This is comparable to the statistical error on the mean redshift per tomographic bin for Stage III surveys (e.g., \citealp{asgari2021}), and below the threshold where cosmological parameter biases become significant ($\Delta z \lesssim 0.04$; \citealp{hildebrandt2016}). The model predictions are within a factor of a few of the requirements for Stage IV surveys ($\Delta z < 0.002(1+z)$; \citealp{mandelbaum2018}), and we note that the Stage IV requirements are contained within the error distribution on the mean bias for VVDS as shown in Figure \ref{fig:VVDS_deltaz}.

We re-iterate that these model predictions are obtained by assuming (fixed) default parameters for the population model described in \S \ref{sec:population}, fitting the uncertainty distribution to the VVDS data to characterize the distribution of photometric errors (as a function of flux), and assuming the zero-point calibration provided by the VVDS collaboration \citep{lefevre2013}.

We note that our data model for VVDS has residual uncertainties that may be responsible for some of the redshift bias. We have assumed standard Johnson $BVRI$ filters, which are approximately correct but there are some differences in detail \citep{lefevre2013}. Zero-point calibration for VVDS was also reported to be challenging, with large (and sometimes differential) zero points required in some bands to achieve consistency between photometry and spectra (Vincent LeBrun, private communication). We have not included calibration uncertainties in our results.
\section{Discussion}
\label{sec:discussion}
While our baseline model is able to accurately recover the tomographic redshift distributions for GAMA and VVDS, a number of improvements are possible. 

The baseline SPS model assumes a simple double power-law star-formation history parameterization. Such a simple SFH parameterization is not expected to capture the full diversity of real star formation histories, and can lead to overly restrictive correlations between important derived quantities (such as SFR and age), which might not be representative of real galaxies. These limitations can be alleviated by non-parametric (binned) star formation history models (e.g., \citealp{leja2019np}), or more physical SFH parameterizations (e.g., \citealp{alarcon2022}).

Regarding the population model, the largest modeling uncertainties are expected to come from the dust attenuation prior, where our baseline model assumed that dust attenuation scales with SFR only. In reality, dust characteristics are expected to be related to the detailed star formation and metallicity enrichment histories, motivating a more sophisticated dust prior model (e.g., \citealp{nagaraj2022}).

The data-model also has a number of simplifying assumptions. SPS modeling errors are expected to vary as a function of rest-frame wavelength, with emission-lines in particular being subject to potentially significant modeling biases \citep{leistedt2022}. In the context of inferring SPS parameters for individual galaxies, photometric error floors have often been used to capture both modeling and calibration uncertainties, in order to increase the uncertainties on inferred SPS model parameters and reduce biases. While this strategy is fine for individual galaxies, increasing variances in order to cover potential biases in this way can lead to over-dispersion in population-level parameters. Therefore, data-modeling efforts should instead focus on parameterizing and modeling biases directly, rather than treating them as extra variance terms. The shape of the noise distribution also merits careful investigation, with (for example) the skewness and tail-weights of photometric measurement errors likely varying between bands, and as a function of flux and background noise levels.

Regarding the selection modeling, so far we have considered the scenario where selection is performed with respect to the measured photometry alone. However, for weak lensing surveys some additional selection cuts will typically be made on the images, such as image quality cuts to ensure reliable shear measurements, image-based star-galaxy separation, deblending, surface-brightness cuts, etc. Because galaxy image characteristics correlate with SPS parameters and redshift, image-based cuts will induce additional selection effects that could modify the resulting redshift distributions.

In Appendix \ref{sec:posterior_derivation} we show that the effect of image-based selection cuts can be addressed by replacing the population model with an effective population prior describing the statistical properties of the galaxy population that passes the image cuts (conditioned on the characteristics of the survey, etc.). Hence, image-based selection can be incorporated into the forward modeling framework by parameterizing the effect of those image cuts on the population prior over SPS parameters and redshift, and inferring those additional hyper-parameters alongside the other population- and data-model parameters. Alternatively, if it can be demonstrated (or orchestrated) that the photometric cuts are sufficiently stronger than any image based cuts, such that the image cuts have a negligible impact on the analysis sample, then those unmodelled selection effects can be safely ignored.

In addition to modeling improvements, inference (or optimization) of population- and data-model parameters from the photometric data should lead to additional improvements in accuracy. For robust inferences, data-model parameters should be self-consistently calibrated using the photometric data themselves, with photometric redshifts expected to be particularly sensitive to zero-point calibrations (but with all data-model parameters playing a role). Regarding the population model, we note that different aspects of the model are better constrained than others by external data. The dust prior and fundamental metallicity relations in particular are expected to be the least well understood and constrained, meriting broader priors on their parameters. The star-forming sequence is somewhat better constrained (e.g., \citealp{leja2021SFS}), while the mass function is relatively tightly constrained (e.g., \citealp{leja2020MF}).
\section{Conclusions}
\label{sec:conclusions}
We have presented a forward modeling framework for photometric surveys, which is capable of accurately predicting the tomographic redshift distributions required for cosmological analyses. Scaling this forward modeling approach to large surveys is made possible by neural emulation of SPS models \citep{alsing2020}.

Forward modeling has a number of advantages over existing methods for estimating cosmological redshift distributions. In contrast to direct-calibration methods, forward modeling does not require external spectroscopic data: it is therefore not hampered by the (lack of) availability of spectroscopic redshifts at the depth required for photometric surveys, and is not vulnerable to biases arising from spectroscopic selection effects that cannot be well-described by re-weighting in broad-band colour-space. In contrast to cross-correlation based estimators, it is not sensitive to galaxy-bias modeling assumptions. Our forward modeling framework also resolves a number of limitations of existing template-based methods, by replacing template-sets with a continuous physical model for galaxy spectra (with associated physical priors), carefully treating selection effects, and enabling self-consistent inference of model parameters describing the galaxy population and data-model.

By explicitly modeling the processes that give rise to the target redshift distributions, forward modeling allows for fine-control over the relevant modeling assumptions. In particular, it creates synergies between galaxy evolution physics and photometric redshift inference: as our constraints on the statistical properties of the galaxy population improve, those lead directly to improved priors on the population model parameters, and hence improved photometric redshift inferences.

We have demonstrated the utility of our forward modeling framework by accurately recovering the redshift distributions for the GAMA and VVDS surveys, validating against their spectroscopic redshifts. The model is able to predict the tomographic redshifts for those two surveys, with biases of $\Delta z\lesssim 0.003$ for GAMA and $\Delta z \simeq 0.01$ for VVDS respectively, without performing inference or optimization of the model parameters describing the galaxy population and photometric calibration. This accuracy is sufficient for ongoing Stage III surveys, and approaching the accuracy requirements of Stage IV surveys. We anticipate that with additional modeling improvements, and optimization of model hyper-parameters, forward modeling can provide a path to accurate cosmological redshift distribution inference for Stage IV surveys.
%

In a companion paper \citep{leistedt2022}, we demonstrate the utility of this forward modeling framework for inferring individual redshifts, including hierarchical calibration of data-model hyper-parameters.

\acknowledgments
\textbf{Author contributions.} 
{\bf JA:} Conceptualization, methodology, software, validation, formal analysis, writing - original draft. {\bf HVP:} Conceptualization, methodology, validation, writing - review \& editing, funding acquisition. {\bf DM:} Conceptualization, methodology, validation, writing - review \& editing, funding acquisition. {\bf JL:} Conceptualization, methodology, validation, data curation, writing - review \& editing.
{\bf BL:} Conceptualization, methodology, writing - review \& editing.

\textbf{Acknowledgements.} We thank George Efstathiou, Angus Wright, Konrad Kuijken, Hendrik Hildebrandt, Will Hartley and Jeff Newman for valuable discussions. We also help Vincent Le Brun and Henry McCracken for helpful communications regarding the VVDS data. This project has received funding from the European Research Council (ERC) under the European Union’s Horizon 2020 research and innovation programme (grant agreement no. 101018897 CosmicExplorer). This work has also been enabled by support from the research project grant ‘Understanding the Dynamic Universe’ funded by the Knut and Alice Wallenberg Foundation under Dnr KAW 2018.0067. JA, HVP and DJM were partially supported by the research project grant “Fundamental Physics from Cosmological Surveys” funded by the Swedish Research Council (VR) under Dnr 2017-04212. The work of HVP was additionally supported by the Göran Gustafsson Foundation for Research in Natural Sciences and Medicine. BL is supported by the Royal Society through a University Research Fellowship.  HVP and DJM acknowledge the hospitality of the Aspen Center for Physics, which is supported by National Science Foundation grant PHY-1607611. The participation of HVP and DJM at the Aspen Center for Physics was supported by the Simons Foundation. 

\appendix

\section{Derivation of the joint posterior including selection effects}
\label{sec:posterior_derivation}
For the purpose of deriving the joint posterior for the forward model described in \S \ref{sec:generate} it is useful to consider the generation of a selected sample directly, as follows (notation is summarized in Table \ref{tab:parameters}).

The total number $N$ of selected galaxies is drawn (assuming Poisson statistics), given the expected number of selected objects under the population model, data model and selection effects:
\begin{align}
    \label{nbar}
    \bar{N}(\Phi_0, \hyper, \nuisance, \observe) = A \int \Phi_0 \rho(z; \hyper) \frac{dV}{dz} P(\sps | z, \hyper)\, P(S | \sps, z, \nuisance, \observe) d\sps\,dz,
\end{align}
where $P(S | \sps, z, \nuisance, \observe)$ is the selection probability for a given set of galaxy parameters (and data-model parameters $\nuisance$ and $\observe$), and $A$ is the survey area. The SPS parameters, redshifts, measurement uncertainties, and data vectors for each (selected) galaxy are then drawn from their respective distributions conditioned on selection. The generative model is hence given by:
\begin{align}
\label{raw_joint_model}
P(\hyper, \Phi_0, & \nuisance, \{\sps, z\}_{1:N}, \{\data,\noise, S\}_{1:N}, N | \mathcal{O}) =  \nonumber \\ 
& P(\hyper) P(\nuisance) P(\Phi_0) \frac{\bar{N}(\Phi_0, \hyper, \nuisance)^Ne^{-\bar{N}(\Phi_0, \hyper, \nuisance)}}{N!}
\,\prod_{i=1}^{N} P(\sps_i, z_i | \hyper, \noise_i, S_i, \nuisance) P(\data_i | \sps_i, z_i, \noise_i, S_i, \nuisance)P(\noise_i|\flux_i, \observe, \hyper, \nuisance, S_i).
\end{align}
Taking a log-uniform prior for the present day volume-density, $P(\Phi_0)= 1/\Phi_0$, and noting that $\bar{N}\propto \Phi_0$, we can marginalize out $\Phi_0$ analytically and obtain:
\begin{align}
\label{phi_marginalized_joint_model}
P(\hyper, \nuisance, \{\sps, z\}_{1:N}, \{\data, \noise, S\}_{1:N}, N | \mathcal{O}) = N^{-1}\;P(\hyper) P(\nuisance) \prod_{i=1}^{N}P(\sps_i, z_i | \hyper, \noise_i, S_i, \nuisance) P(\data_i | \sps_i, z_i, \noise_i, S_i, \nuisance) P(\noise_i|\flux_i, \observe, \hyper, \nuisance, S_i).
\end{align}
The joint posterior is hence given by:
\begin{align}
\label{joint-posterior-selection}
P(\hyper, \nuisance, & \{\sps, z\}_{1:N}| \{\data, \noise, S\}_{1:N}, N, \mathcal{O}) = P(\hyper) P(\nuisance) \prod_{i=1}^{N}P(\sps_i, z_i | \hyper, \noise_i, S_i, \nuisance) P(\data_i | \sps_i, z_i, \noise_i, S_i, \nuisance),
\end{align}
where we have dropped the $P(\noise_i|\flux_i, \observe, \hyper, \nuisance, S_i)$ term in the posterior, since its sensitivity to the latent and hyper-parameters is typically negligible compared to the likelihood and prior terms.

Written this way, the population model and likelihood terms enter conditioned on selection. This parameterization has the drawback that the population model conditioned on selection is typically hard to parameterize directly, and becomes a survey-specific quantity. Also, if selection cuts involve more than simple (independent) flux or S/N cuts in each band, then the data-model conditioned on selection is also hard to compute directly.

It is instead desirable to re-write the model in terms of the population model for the background galaxy population, and the likelihood without selection. Using the chain rule, the population model and likelihood terms can be re-written as:
\begin{align}
    P(\sps, z | \hyper, \noise, S, \nuisance) &= \frac{P(\sps, z | \hyper)P(S | \sps, z, \noise, \nuisance)}{P(S|\hyper, \noise, \nuisance)}, \nonumber \\
    P(\data | \sps, z, \noise, S, \nuisance) &= \frac{P(\data |\sps, z,  \noise, \nuisance)P(S|\data, \noise)}{P(S | \sps, z, \noise, \nuisance)}.
\end{align}
Inserting these into Equation \eqref{joint-posterior-selection}, we obtain (after cancellation and dropping parameter-independent terms):
\begin{align}
\label{joint_posterior_reparam}
P(\hyper, \nuisance, \{\sps, z\}_{1:N} | \{\data, \noise, S\}_{1:N}, N, \mathcal{O}) = P(\hyper) P(\nuisance)\times \prod_{i=1}^{N}\frac{P(\sps_i, z_i | \hyper)P(\data_i | \sps_i, z_i, \noise_i, \nuisance)}{P(S_i | \hyper, \noise_i, \nuisance)},
\end{align}
where the selection term in the denominator is given by:
\begin{align}
    P(S | \hyper, \noise, \nuisance) = \int P(S | \data, \noise) P(\data | \sps, z, \noise, \nuisance) P(\sps, z | \hyper)\,d\data \,d\sps \,dz.
\end{align}
Note that selection term only depends explicitly on the hyper- and data-model parameters $\hyper$ and $\nuisance$. In the special case where one is only interested in inferring the latent parameters $(\sps, z)$ for each galaxy with the hyper- and data-model parameters fixed, selection appears to ``drop out" of the problem and one should infer the latent parameters under the population prior and data models without selection. Note that in this case (although it might seem counter-intuitive) selection effects are still included properly: they only enter implicitly via the ensemble of selected galaxies that made it into the analysis sample.

When inferring hyper- and data-model parameters though, the selection term is important: the high-dimensional integral over parameter and data space in Equation \eqref{pdet} usually represents the computational bottleneck for sampling Bayesian hierarchical models under selection effects.
\subsection{Selection effects on both photometry and images}
In the model derived above, we assumed that selection was performed with respect to the photometric data vector only (i.e., based on the measured fluxes and their uncertainties). In reality, for weak lensing surveys some selection cuts will occur at the level of images, such as image quality cuts to insure robust shear measurements, removal of blended objects, surface brightness cuts, etc. Therefore, it is useful to consider the impact of image-based selection cuts on the generative model. To this end, in this section we derive the joint generative model for galaxy photometry and images, and subsequently marginalize over the images to explore the typical case where one wants to perform redshift inference with respect to photometry only, but needs to account for image-based selection cuts.

We will assume that galaxy images are characterized by parameters $\appearance$ (in addition to the SPS parameters and redshift). We denote galaxy image data-vectors and uncertainties by $\Data$ and $\Noise$ respectively, where $\Data$ can be taken to mean either the full pixelized image, or some low-dimensional summary statistics derived from the galaxy images on which selection is performed. For the purpose of this derivation, we distinguish photometric and image-based selection cuts by $S_d$ and $S_D$ respectively, and use $S$ to denote combined selection.

Conceptually, the forward model proceeds as before. The total number $N$ of selected galaxies is drawn (assuming Poisson statistics), given the expected number of selected objects:
\begin{align}
    \label{nbar}
    \bar{N}(\Phi_0, \hyper, \nuisance, \observe) = A \int \Phi_0 \rho(z; \hyper) \frac{dV}{dz} P(\appearance, \sps | z, \hyper)  P(S | \appearance, \sps, z, \nuisance, \observe) d\appearance d\sps dz,
\end{align}
The parameters, redshifts, measurement uncertainties, and data vectors for each (selected) galaxy are then drawn from their respective sampling distributions, conditioned on selection. The generative model is hence given by:
\begin{align}
\label{raw_joint_model}
P(\hyper, & \Phi_0, \nuisance, \{\sps, \appearance, z\}_{1:N}, \{\data, \Data, \noise, \Noise, S\}_{1:N}, N | \mathcal{O}) = P(\hyper) P(\nuisance) P(\Phi_0) \frac{\bar{N}(\Phi_0, \hyper, \nuisance)^Ne^{-\bar{N}(\Phi_0, \hyper, \nuisance)}}{N!} \nonumber \\
&\times \prod_{i=1}^{N}P(\sps_i, \appearance_i, z_i | \hyper, \noise_i, \Noise_i, S_i, \nuisance) \,P(\data_i | \sps_i, z_i, \noise_i, S_{d,i}, \nuisance) P(\Data_i | \appearance_i, \Noise_i, S_{D,i}, \nuisance) \,P(\noise_i|\flux_i, \observe, \hyper, \nuisance, S_i) P(\Noise_i|\observe, \hyper, \nuisance, S_i),
\end{align}
where we have assumed that the errors on the photometry and images are uncorrelated. We note that while it is true that the fluxes are also summary statistics extracted from the images, other summary statistics $\Data$ extracted from the images on which cuts are made (e.g., image quality flags) are likely to characterize very different features of the raw pixelized galaxy image, so the assumption that their errors are uncorrelated is probably reasonable. 

Taking a log-uniform prior for the present day volume-density and marginalizing out $\Phi_0$ gives:
\begin{align}
\label{phi_marginalized_joint_model}
P(\hyper, \nuisance, \{\sps, \appearance, z\}_{1:N}, \{\data, \Data, \noise, \Noise, S\}_{1:N}, N | \mathcal{O}) = N^{-1} \,P(\hyper) & P(\nuisance) \prod_{i=1}^{N}P(\sps_i, \appearance_i, z_i | \hyper, \noise_i, \Noise_i, S_i, \nuisance) P(\data_i | \sps_i, z_i, \noise_i, S_{d,i}, \nuisance) \nonumber \\
& \times P(\Data_i | \appearance_i, \Noise_i, S_{D,i}, \nuisance) P(\noise_i|\flux_i, \observe, \hyper, \nuisance, S_i) P(\Noise_i|\observe, \hyper, \nuisance, S_i).
\end{align}
Using the chain rule, the population model and likelihood terms can be re-written as:
\begin{align}
    P(\sps, \appearance, z | \hyper, \noise, \Noise, S, \nuisance) &= \frac{P(\sps, \appearance, z | \hyper)P(S | \sps, \appearance, z, \noise, \Noise, \nuisance)}{P(S|\hyper, \noise, \Noise, \nuisance)}, \nonumber \\
    P(\data | \sps, z, \noise, S_d, \nuisance) &= \frac{P(\data |\sps, z,  \noise, \nuisance)P(S_d|\data, \noise)}{P(S_d | \sps, z, \noise, \nuisance)} \nonumber \\
    P(\Data | \appearance, \Noise, S_D, \nuisance) &= \frac{P(\Data |\appearance, \Noise, \nuisance)P(S_D | \Data, \Noise)}{P(S_D|\appearance, \Noise, \nuisance)}.
\end{align}
Inserting these into Equation \eqref{phi_marginalized_joint_model}, we obtain (after cancellation):
\begin{align}
\label{reparameterized_joint_model}
P(\hyper, \nuisance, \{\sps, \appearance, z\}_{1:N}, \{\data, \Data, \noise, \Noise, S\}_{1:N}, N | \mathcal{O}) = N^{-1}& P(\hyper) P(\nuisance)\prod_{i=1}^{N} P(\sps_i, \appearance_i, z_i | \hyper) \, P(\data_i | \sps_i, z_i, \noise_i, \nuisance) P(\Data_i | \appearance_i, \Noise_i, \nuisance) \nonumber \\
&\times \frac{P(S_{d,i}| \data_i, \noise_i)P(S_{D,i} | \Data_i, \Noise_i)}{P(S_i | \hyper, \noise_i, \Noise_i, \nuisance)} P(\noise_i|\observe, \hyper, \nuisance, S_i) P(\Noise_i|\observe, \hyper, \nuisance, S_i).
\end{align}
For joint modeling of both photometry and images, the joint posterior is hence given by:
\begin{align}
\label{joint_posterior_phot_image}
P(\hyper, \nuisance, \{\sps, \appearance, z\}_{1:N} |& \{\data, \Data, \noise, \Noise, S\}_{1:N}, N, \mathcal{O}) = \nonumber \\
& P(\hyper) P(\nuisance) \prod_{i=1}^{N}P(\sps_i, \appearance_i, z_i | \hyper) P(\data_i | \sps_i, z_i, \noise_i, \nuisance)\, P(\Data_i | \appearance_i, \Noise_i, \nuisance) \times \frac{1}{P(S_i | \hyper, \noise_i, \Noise_i, \nuisance)},
\end{align}
where the selection term in the denominator is given by:
\begin{align}
\label{pdet_phot_image}
    P(S | \hyper, \noise, \Noise, \nuisance) = \int P(S_D | \Data,\Noise) P(S_d | \data, \noise) P(\data | \sps, z, \noise, \nuisance) P(\Data|\appearance, \Noise, \nuisance) P(\sps, \appearance, z | \hyper)\,d\data \,d\Data \,d\sps \,d\appearance \,dz,
\end{align}
and as before, we have dropped the $P(\noise_i|\observe, \hyper, \nuisance, S_i)$ and $P(\Noise_i|\observe, \hyper, \nuisance, S_i)$ terms, based on the notion that their parameter sensitivity will be negligible compared to the likelihood and prior terms.

Now, in order to determine the effect of image-based cuts when performing inference with respect to photometry alone, we need to marginalize over the image parameters, image data vectors, and their uncertainties. Taking the joint generative model in Equation \eqref{reparameterized_joint_model} and marginalizing over the image data vectors $\Data_{1:N}$, one obtains:
\begin{align}
\label{image_marginalized_joint_model}
P(\hyper, \nuisance, \{\sps, \appearance, z\}_{1:N}, \{\data, \noise, \Noise, S\}_{1:N}, N | \observe) = N^{-1}& P(\hyper) P(\nuisance) \prod_{i=1}^{N}P(\sps_i, \appearance_i, z_i | \hyper)P(\data_i | \sps_i, z_i, \noise_i, \nuisance) \nonumber \\
& \times\frac{P(S_{d,i}| \data_i, \noise_i)P(S_D |\appearance_i, \Noise_i, \nuisance)}{P(S_i | \hyper, \noise_i, \Noise_i, \nuisance)} P(\noise_i|\flux_i, \observe, \hyper, \nuisance, S_i)P(\Noise_i|\observe, \hyper, \nuisance, S_i).
\end{align}
We can then re-write the selection term in the denominator as,
\begin{align}
P(S | \hyper, \noise, \Noise, \nuisance) = P(S_d | S_D, \hyper, \noise, \nuisance) P(S_D | \hyper, \Noise, \nuisance),
\end{align}
and absorb $P(S_D | \hyper, \Noise, \nuisance)$ and $P(S_D |\appearance, \Noise, \nuisance)$ into the population prior term (again using the chain rule) to give:
\begin{align}
\label{reparameterized_image_marginalized_joint_model}
P(\hyper, \nuisance, \{\sps, \appearance, z\}_{1:N}, \{\data, \noise, \Noise, S\}_{1:N}, N | \observe) = N^{-1}& P(\hyper)  P(\nuisance) \prod_{i=1}^{N}P(\sps_i, \appearance_i, z_i | \hyper, S_{D,i}, \Noise_i, \nuisance) P(\data_i | \sps_i, z_i, \noise_i, \nuisance) \nonumber \\
& \times \frac{P(S_{d,i}| \data_i, \noise_i)}{P(S_{d,i} | S_{D,i}, \hyper, \noise_i, \nuisance)} P(\noise_i|\flux_i, \observe, \hyper, \nuisance, S_{d,i})P(\Noise_i|\observe, \hyper, \nuisance, S_{D,i}).
\end{align}
Marginalizing over the image uncertainties $\Noise_{1:N}$ and parameters governing the galaxy images $\appearance_{1:N}$ then gives:
\begin{align}
\label{reparameterized_fully_marginalized_joint_model}
P(\hyper, \nuisance, & \{\sps, z\}_{1:N}, \{\data, \noise, S\}_{1:N}, N | \observe) = \nonumber \\
& N^{-1}P(\hyper) P(\nuisance) \prod_{i=1}^{N}P(\sps_i, z_i | \hyper, S_{D,i}, \nuisance, \observe) P(\data_i | \sps_i, z_i, \noise_i, \nuisance) \,
\frac{P(S_{d,i}| \data_i, \noise_i)}{P(S_{d,i} | S_{D,i}, \hyper, \noise_i, \nuisance)} P(\noise_i|\flux_i, \observe, \hyper, \nuisance, S_{d,i}).
\end{align}
Dropping parameter-independent terms (and assuming that the parameter sensitivity of $P(\noise_i|\flux_i, \observe, \hyper, \nuisance, S_{d,i})$ is negligible relative to the prior and likelihood terms) then gives the posterior:
\begin{align}
\label{reparameterized_fully_marginalized_joint_posterior}
P(\hyper, \nuisance, \{\sps, z\}_{1:N}, | \{\data, \noise, S\}_{1:N}, N, \observe) \propto P(\hyper) P(\nuisance) \prod_{i=1}^{N}P(\sps_i, z_i | \hyper, S_D, \nuisance, \observe) P(\data_i | \sps_i, z_i, \noise_i, \nuisance) \frac{1}{P(S_d | S_D, \hyper, \noise_i, \nuisance)},
\end{align}
where the selection term in the denominator is given by:
\begin{align}
\label{pdet_II}
    P(S_d | S_D, \hyper, \noise, \nuisance) = \int  P(S_d | \data, \noise) P(\data | \sps, z, \noise, \nuisance) P(\sps, \appearance, z | \hyper, S_D, \nuisance, \observe)\,d\data\, d\sps\, dz.
\end{align}
Crucially, note how this posterior has exactly the same form as in Equation \eqref{joint_posterior_reparam}, but the image-based selection cuts have been completely absorbed into an effective population prior model for galaxies that pass the image-based selection cuts. Those selection effects will not be well-specified without assuming a detailed model for the joint distribution of SPS parameters and galaxy images; therefore, they will typically need parameterizing and inferring.
\section{Star forming sequence model}
\label{sec:sfs_surrogate}
\citet{leja2021SFS} use a normalizing flow to model the star-forming sequence (SFS), i.e., the distribution of star-formation rate conditioned on mass and redshift. While their flow provides a state-of-the-art measurement of the SFS, their model utilized a dummy variable that needs integrating over in order to compute the log-probability $P(\mathrm{SFR} | M, z)$, and has no explicit constraints to ensure that it extrapolates sensibly below the mass-complete limit for the data that it was fitted to. To address these two issues, we construct a surrogate model to approximate their normalizing flow. The SFS is modeled as a mixture of a Gaussian (characterizing star forming galaxies) and a SinhArcSinh distribution, characterizing quiescent galaxies:
\begin{align}
    P(\log10\mathrm{SFS} | \mathcal{M}, z) = r_\mathrm{q}(\mathcal{M}, z) \mathcal{S}(\log10\mathrm{SFS} | \mu_\mathrm{q}(\mathcal{M}, z), \sigma_\mathrm{q}(\mathcal{M}, z), k_\mathrm{q}) + (1-r_\mathrm{q}(\mathcal{M}, z)) \, \mathcal{N}(\log10\mathrm{SFS} | \mu_\mathrm{sf}(\mathcal{M}, z), \sigma_\mathrm{sf}),
\end{align}
where $\mathcal{M}\equiv \log10 M$, $\mathcal{N}$ denotes the normal distribution and $\mathcal{S}$ denotes the SinhArcSinh distribution, defined as a bijection of the unit normal distribution: $x\rightarrow \mu + \sigma\, \mathrm{sinh}^{-1}(\mathrm{sinh}(x) + k)$.

The location, scale of the SinhArcSinh, the relative weight of the two mixture components, and the location Gaussian, are functions of mass and redshift defined by:
\begin{align}
    \mu_\mathrm{sf}(\mathcal{M}, z) & = \left[ a(z)\Theta(\mathcal{M} - \mathcal{M}_\mathrm{sf}^*(z)) + b(z)\Theta(\mathcal{M}_\mathrm{sf}^*(z) - \mathcal{M})\right] (\mathcal{M} - \mathcal{M}_\mathrm{sf}^*(z)) + c(z) \nonumber \\
    \mu_\mathrm{q}(\mathcal{M}, z) & = \left[ d(z)\Theta(\mathcal{M} - \mathcal{M}_\mathrm{sf}^*(z)) + e(z)\Theta(\mathcal{M}_\mathrm{sf}^*(z) - \mathcal{M})\right] (\mathcal{M} - \mathcal{M}_\mathrm{sf}^*(z)) + f(z) - 1 \nonumber \\
    \sigma_\mathrm{q}(\mathcal{M}, z) & = \sigma_{q0} + (\sigma_{q0} - \sigma_{q1}) \mathrm{sigmoid}((\mathcal{M} - \sigma_{q2}) / \sigma_{q3}) \nonumber \\
    r_\mathrm{q}(\mathcal{M}, z) & = \mathrm{sigmoid}(j) * \mathrm{sigmoid}((\mathcal{M} - (g + h*z)) / i )
\end{align}
where the functions $a(z)$ through $f(z)$, as well as $\mathcal{M}_\mathrm{sf}^*(z)$ and $\mathcal{M}_\mathrm{q}^*(z)$, are quadratic functions in redshift, defined as $a(z) = a_0 + a_1 z + a_2 z^2$ etc. 

We fit this parametric form to the normalizing flow of \citet{leja2021SFS} by generating $10^6$ samples from the flow, and maximizing the total log likelihood of those samples under our parametric model. Optimization is performed with \texttt{Adam} \citep{kingma2014adam}, with a learning rate of $10^{-4}$, no mini-batching and $10^4$ epochs. Fitted parameters are given in Table \ref{tab:sfs_fit}.
\begin{table}
\centering
\begin{tabularx}{0.6\textwidth}{cc}
\toprule
Parameter & Fitted value \tabularnewline
\hline \tabularnewline
$(a_0, a_1, a_2)$ & $(-0.15040097,  0.9800668 , -0.50802046)$ \tabularnewline
$(b_0, b_1, b_2)$ & $(1.0515388 , -0.28611764,
        0.02131329)$ \tabularnewline
$(c_0, c_1, c_2)$ & $(0.05053138,  1.0766244 , -0.02015052)$ \tabularnewline
$(d_0, d_1, d_2)$ & $(-0.13125503,
        0.7205097 , -0.18212801)$ \tabularnewline
$(e_0, e_1, e_2)$ & $(1.5429502 , -1.5872463 , -0.04843145)$ \tabularnewline
$(f_0, f_1, f_2)$ & $(0.65359867,  0.92735046, -0.17695354)$ \tabularnewline
$(g, h, i, j)$ & $(10.442122  ,  0.56389964,
        0.7500511 ,  2.0604856)$ \tabularnewline
$(\mathcal{M}_\mathrm{q0}^*, \mathcal{M}_\mathrm{q1}^*, \mathcal{M}_\mathrm{q2}^*)$ & $(10.611108  ,  0.08009169, -0.06575003)$ \tabularnewline
$(\mathcal{M}_\mathrm{sf0}^*, \mathcal{M}_\mathrm{sf1}^*, \mathcal{M}_\mathrm{sf2}^*)$ & $(10.335057  , -0.3050156 ,  0.5491848)$ \tabularnewline
$(\sigma_{q0}, \sigma_{q1}, \sigma_{q2}, \sigma_{q3})$ & $(0.54855245, 0.44964817, 11.159543  ,  0.11614972)$ \tabularnewline
$\sigma_\mathrm{sf}$ & $0.3912887$ \tabularnewline
$k_\mathrm{q}$ & $-1.5658572$ \tabularnewline

\tabularnewline

\hline
\end{tabularx}
\caption{Fitted values of the surrogate SFS model.}
\label{tab:sfs_fit}
\end{table}
\section{Setting priors on derived quantities using normalizing flows}
\label{sec:nf_priors}
As described in \ref{sec:population}, we want to put a prior on the SFH parameters such that the resulting prior on the SFR is given by our desired assumptions about the star-forming sequence (SFS). To re-cap from the main text, we construct a prior over SFH parameters as:
\begin{align}
    P(\alpha, \beta, \tau | M, z) = \frac{\pi_0(\alpha, \beta, \tau)\;P(\mathrm{SFR} | M, z)}{P_0(\mathrm{SFR} | z)}
\end{align}
where $\pi_0(\alpha, \beta, \tau)$ is the baseline (uniform) prior on the SFH parameters, $P(\mathrm{SFR} | M, z)$ is the target SFS prior, and $P_0(\mathrm{SFR} | z)$ is the implicit prior on SFR implied by $\pi_0$. The implicit SFR prior is defined by the surface integral
\begin{align}
P_0(\mathrm{SFR} | z) = \int_{\mathrm{SFR}=\mathrm{const.}} \pi_0(\alpha, \beta, \tau) dS,
\end{align}
where $dS$ is the surface element in the SFH parameter space $(\alpha, \beta, \tau)$. In order to circumvent having to compute those surface integrals directly every time we need to evaluate the prior density, we train a normalizing flow to learn $P_0(\mathrm{SFR} | z)$ as follows. We construct a training set by drawing SFH parameters from the baseline prior $\pi_0(\alpha, \beta, \tau)$, which we take as log-uniform in $\alpha$ and $\beta$, and uniform in $\tau$, over the prior ranges given in Table \ref{tab:sps_parameters}, and redshifts drawn uniformly between $0$ and $2$. For each baseline prior (and redshift) sample, we compute the specific SFR for those SFH and redshift parameters, defined as the fractional mass formed in the last 100$\mathrm{Myr}$. This provides a training set of $\{\mathrm{sSFR}, \alpha, \beta, \tau, z\}$, on which we can train a conditional density estimator to learn $P_0(\mathrm{sSFR} | z)$.

\begin{figure}
\centering
\includegraphics[width = 18cm]{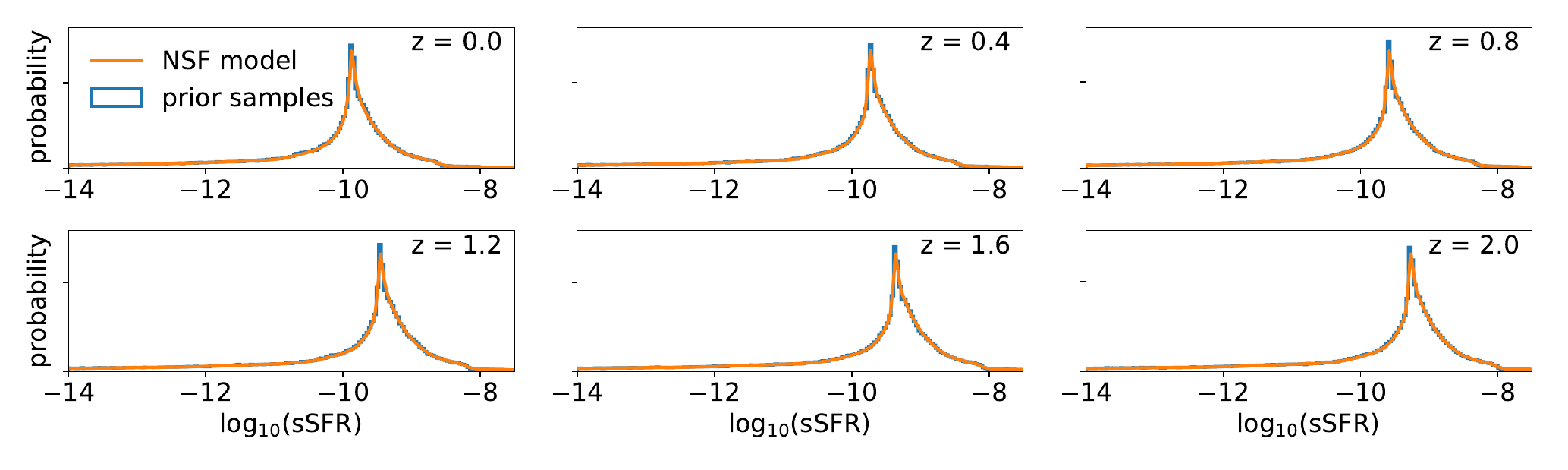}
\caption{Samples from the prior on the star formation rate implied by taking uniform priors over the double power-law SFH parameters $(\log10 \alpha, \log10 \beta, \tau)$ over the ranges specified in Table \ref{tab:sps_parameters} is shown as blue histograms, while the learned normalizing flow model for the implied distribution (as a function of redshift) is shown in orange.}
\label{fig:sfs_calibrator}
\end{figure}

We train a neural spline flow \citep{durkan2019} to learn $P_0(\log10\mathrm{sSFR} | z)$, with $16$ spline knots spaced between $-15$ and $-6$ (in $\log10\mathrm{sSFR}$), a single hidden layer with $16$ units and leaky-ReLU activation functions, and a base Gaussian density with location $-11$ and scale $1$. Training is performed using \texttt{Adam} with a learning rate of $1e-2$ for $1000$ epochs, with no batching on a training set of $10^6$ samples (generated as described above). The implicit prior $P_0(\log10\mathrm{sSFR} | z)$ and the trained normalizing flow is shown in Figure \ref{fig:sfs_calibrator}, and use of the trained normalizing flow to divide out the implicit SFR prior is shown in Figure \ref{fig:sfs_calibrated}.
\bibliography{spsbhm}

\end{document}